\documentclass[aps,prb,twocolumn,showpacs,superscriptaddress,floatfix]{revtex4-1}

\pdfoutput=1

\usepackage{amsmath,amssymb}
\usepackage{graphicx}
\usepackage{color}
\usepackage{rotating}
\usepackage{bm}
\usepackage{setspace}
\usepackage{hyperref}

\hypersetup{
colorlinks=true,
urlcolor= blue,
citecolor=blue,
linkcolor= blue,
bookmarks=true,
bookmarksopen=false,
}

\renewcommand{\vec}[1]{\bm{#1}}

\begin{document}

\preprint{APS/123-QED}

\title{\textit{Ab-initio} Tight-Binding Hamiltonian for Transition Metal Dichalcogenides}


\author{Shiang Fang}
\affiliation{Department of Physics, Harvard University, Cambridge, Massachusetts 02138, USA.}
\author{Rodrick Kuate Defo}
\affiliation{Department of Physics, Harvard University, Cambridge, Massachusetts 02138, USA.}
\author{Sharmila N. Shirodkar}
\affiliation{John A. Paulson School of Engineering and Applied Sciences, Harvard University, Cambridge, Massachusetts 02138, USA.}
\author{Simon Lieu}
\affiliation{Department of Physics, Harvard University, Cambridge, Massachusetts 02138, USA.}
\author{Georgios A. Tritsaris}
\affiliation{John A. Paulson School of Engineering and Applied Sciences, Harvard University, Cambridge, Massachusetts 02138, USA.}
\author{Efthimios Kaxiras}
\affiliation{Department of Physics, Harvard University, Cambridge, Massachusetts 02138, USA.}
\affiliation{John A. Paulson School of Engineering and Applied Sciences, Harvard University, Cambridge, Massachusetts 02138, USA.}

\date{\today}

\begin{abstract}
We present an accurate \textit{ab-initio} tight-binding hamiltonian for the transition-metal dichalcogenides, MoS$_2$, MoSe$_2$, WS$_2$, WSe$_2$, with a minimal basis (the \textit{d} orbitals for the metal atoms and \textit{p} orbitals for the chalcogen atoms) based on a transformation of the Kohn-Sham density function theory (DFT) hamiltonian to a basis of maximally localized Wannier functions (MLWF). The truncated tight-binding hamiltonian (TBH), with only on-site, first and partial second neighbor interactions, including spin-orbit coupling, provides a simple physical picture and the symmetry of the main band-structure features. Interlayer interactions between adjacent layers are modeled by transferable hopping terms between the chalcogen \textit{p} orbitals. The full-range tight-binding hamiltonian (FTBH) can be reduced to hybrid-orbital k $\cdot$ p effective hamiltonians near the band extrema that captures important low-energy excitations. These \textit{ab-initio} hamiltonians can serve as the starting point for applications to interacting many-body physics including optical transitions and Berry curvature of bands, of which we give some examples.
\end{abstract}

\pacs{}
\maketitle


\section{\label{sec:level1}INTRODUCTION}
The successful isolation of a single atomic layer of graphene\cite{graphene_2004} ushered in a new era in the study of two-dimensional materials, but its gapless band structure has hindered its applications in electronic devices that depend on the presence of a band gap. In contrast to graphene, layered transition metal dichalcogenides (TMDCs)\cite{2dmaterial_rev1,2dmaterial_rev2,2dmaterial_rev3,mos2_transistor} with chemical formula MX$_2$ (representative examples being M=Mo,W and X=S,Se), are semiconductors with direct or indirect band-gaps that depend on the number of layers and are in the range of visible light (1-2 eV). These materials also display a whole spectrum of phenomena such as superconductivity\cite{TMDC_SC}, magnetism\cite{TMDC_magnetism}, charge density waves (in TaS$_2$) observed by experiment\cite{tas2_cdw1,tas2_cdw2} and topological insulator phases predicted by theory\cite{TMDC_TI1,TMDC_TI2}. For monolayer TMDCs, the broken inversion symmetry enables optical control of valley degrees of freedom\cite{MoS2_opt_polarization} and the strong spin-orbit coupling leads to valley-spin coupling\cite{tmdc_spin_valley}. The two valleys of the $k$-points labeled K$_\pm$ can be manipulated by breaking the time-reversal symmetry with the optical Stark effect\cite{valley_break_opt}, by an external magnetic field\cite{valley_break_magf1,valley_break_magf2,valley_break_magf3,layer_pseudospin} or by a magnetic substrate\cite{valley_break_mags}.  Van der Waals heterostructures\cite{2dmaterial_rev1, het_structure} with TMDCs or other two-dimensional layered materials such as hBN (an insulator) can also be fabricated. In heterostructures with incommensurate lattice periodicity, the induced interlayer potential from Moir\'{e} patterns provides an additional way to control electronic properties\cite{het_moire1,het_moire2}. With recent advances in the experimental techniques for synthesizing these materials, novel optoelectronic, valleytronic and spintronic\cite{tmdc_spin_valley} applications have been proposed, that take advantage of the interplay between different features and the spin, valley and layer degrees of freedom.


To address the whole range of interesting phenomena in the TMDC systems, it is crucial to have a simple and accurate model as a guide to the physics and the symmetries involved. Thus far, only two types of approaches have been considered toward this goal: i) oversimplified models such as k $\cdot$ p theory and tight-binding hamiltonians with few bands involving only the $d$ orbitals of the metal atoms\cite{tmdc_spin_valley,mos2_tb1,mos2_tb2,tmdc_kp}; ii) fitted models that contain all the details of the \textit{ab-initio} calculations, often parametrized as a tight-binding hamiltonian with a large number of parameters fitted to reproduce the \textit{ab-initio} results\cite{mos2_tb_dft1,mos2_tb_dft2}. While both types of approaches are valuable and interesting, they each have certain deficiencies: In the first type of approach, the physics is transparent due to the small number of basis functions and parameters involved, but not necessarily accurate or reliable.  In the second type of approach, the fitted models have the required precision, but lack a simple basis and transparency in the physics; the large number of parameters involved in the fitting, which is not unique, may also lead to inconsistent results especially in relation to surface states as explained below.

In this work, we present simplified models for TMDCs based on the full-range eleven-band tight-binding hamiltonian (FTBH) obtained by Wannier transformation of density functional theory (DFT) results\cite{mos2_tb_wan}.  The truncated tight-binding hamiltonian (TBH) that retains only the first and partial second neighbor coupling is a good compromise, with little sacrifice of accuracy, while it yields clear interpretation of the physics in terms of the simple basis involved: (1) it contains all relevant orbitals near the Fermi level, which are \textit{p} orbitals from X (chalcogen) atoms and \textit{d} orbitals from M (metal) atoms; (2) it preserves the phase and orbital information and character as well as all the crystal symmetries; (3) it preserves the orthogonality of basis functions which is crucial in constructing interacting many-body theories; (4) it provides a systematic way of introducing higher order corrections and spin-orbit coupling terms; (5) it allows for interlayer hopping terms; (6) it allows the calculation of optical transitions and the Berry curvature of bands; (7) most importantly, it contains no empirical parameters and is therefore a true \textit{ab-initio} tight-binding hamiltonian. We also provide a simple recipe, based on correcting the band gap and band width, to show that scaling of the parameters derived from DFT results gives a good representation of the more accurate GW calculations. We only give these results for one case, MoS2, due to the considerable computational cost of obtaining GW band structure results, but expect the procedure to be applicable to other similar materials. To augment the accuracy of the TBH, we also construct k $\cdot$ p hamiltonians for the low-energy bands around the $\Gamma$ and K points of the Brillouin Zone (BZ) by expanding the FTBH.

Building on the single-layer TBH and properly symmetrized orbitals, we derive a set of hopping matrix elements that accurately describe interlayer coupling within the tight-binding approximation. These matrix elements cover a range of possible distance between atomic sites in different layers, obtained by sliding two layers relative to each other.  Surprisingly, the dependence of interlayer coupling parameters on distance shows very simple scaling with distance and is {\em independent} of angular dependence.  This makes possible the calculation of the electronic properties of configurations that involve arbitrary twists between successive layers, as well as calculation of inter-band optical transitions and Berry curvature of heterostructures consisting of different combinations of layers. Thus, our tight-binding scheme can be used to search efficiently for interesting behavior in a wide range of layer arrangements.

The paper is organized as follows: In Sec. II we introduce the DFT and Wannier formalism as the tools employed in studying the electronic structure of the relaxed TMDC crystal structures which are defined in Sec. III. To construct the TBH, we start with a monolayer unit in Sec. IV, where we discuss in detail the symmetries, spin-orbit coupling terms and the comparison to DFT results. Based on the hamiltonian for the monolayer unit, we present in Sec. V the generalization to multiple layers, modeled by introducing interlayer coupling through a transferable interaction term. In Sec. VI, we derive the complementary k $\cdot$ p hamiltonians, based on the relevant orbitals identified from the preceding analysis, that capture all the low-energy physics of these materials. The applications for optical absorption and Berry phase with the TBH is discussed in Sec. VII. In Sec. VIII, we give a summary of the models and their potential applications. The reduction of the TBH parameter space from symmetry considerations, the numerical values of the parameters, and the GW calculations are presented in the Appendix.

\section{\label{sec:level1}NUMERICAL METHODS: DFT AND WANNIER FORMALISM}

We perform the DFT calculations using the Vienna Ab initio Simulation Package (VASP)\cite{vasp1,vasp2}. The interaction between ionic cores and valence electrons is described by pseudo-potentials of the Projector Augmented-Wave (PAW) type. The exchange-correlation energy of electrons is treated within the Generalized Gradient Approximation (GGA) as parametrized by Perdew, Burke and Ernzerhof (PBE)\cite{pbe}. We performed calculations with and without spin-orbit coupling to guide the construction of the corresponding correction terms in the tight-binding hamiltonian. A slab geometry is employed to model single or double layers with a 20 \AA $ $ vacuum region between periodic images to minimize the interaction between slabs. The crystal structure is relaxed until the Hellmann-Feynman forces are smaller in magnitude than 0.01eV/\AA $ $ for each atom. The plane-wave energy cutoff we use is $450$ eV with a reciprocal space grid of size 25 $\times$ 25 $\times$ 1. Due to the underestimation of band-gaps and band widths in DFT, we also performed GW quasi-particle calculations to correct the MoS$_2$ DFT results.

An alternative way to represent the DFT or GW band structure is to transform the Bloch basis into a basis of maximally-localized Wannier functions (MLWF)\cite{mlwf} as implemented in the Wannier90 code. To perform this Wannier unitary transformation for TMDCs, we use the seven highest valence bands and four lowest conduction bands composed of \textit{d} orbitals for the metal atoms and \textit{p} orbitals for the chalcogen atoms to form the localized Wannier functions. The initial projections are chosen to be the atomic \textit{p}/\textit{d} orbitals and the final converged Wannier functions are close to the localized atomic orbitals. The effective hamiltonian in the Wannier basis is interpreted as the full-range \textit{ab-initio} tight-binding hamiltonian (FTBH). With a 25 $\times$ 25 $\times$ 1 $k$-point grid sampling, the numerical accuracy of the FTBH is usually within a few meV compared to DFT or GW bands. Based on this FTBH, the truncated tight-binding hamiltonian (TBH) that retains only first and partial second neighbor couplings is often adequate to elucidate the nature of the bands. By adding more terms beyond nearest neighbor couplings, the numerical accuracy can be improved until it reaches DFT or GW-level results. This provides a systematic way to improve the tight-binding hamiltonian.

We wish to reiterate the importance of properly derived TBH parameters: the parameters obtained in the way discussed here are solely determined by the overlap integrals for orbitals and the matrix elements from the full \textit{ab-initio} calculation. The alternative approach in constructing a TBH often consists of optimizing the model parameters by fitting the band energies of either DFT calculations or experimental results. In that approach, the orbital character of the bands or the phases of the overlap integral are not explicitly considered and maybe subject to overfitting. For example, in the nearest-neighbor tight-binding hamiltonian of single layer graphene, the band structure is invariant under a sign change of the nearest hopping parameter $t$. However, the sign of this hopping parameter can be determined from photoemission experiment\cite{graphene_phase}. We find that the sign we obtain from our procedure for constructing the TBH is consistent with Wannier analysis but is not determined from a blind fitting procedure. In some cases the discrepancy due to overfitting can be even more subtle. As another example, a study related to Bi-Sb alloy\cite{bi_sb_surface} pointed out that it is possible to arrive at fitted hamiltonians that are inconsistent in the description of surface states, which was attributed to the mistaken mirror Chern number. For topological materials, it is even more crucial to get the signs of the parameters right in order to give the correct topological invariants and Berry curvature.

\begin{figure}
\centering
\includegraphics[width=0.45\textwidth]{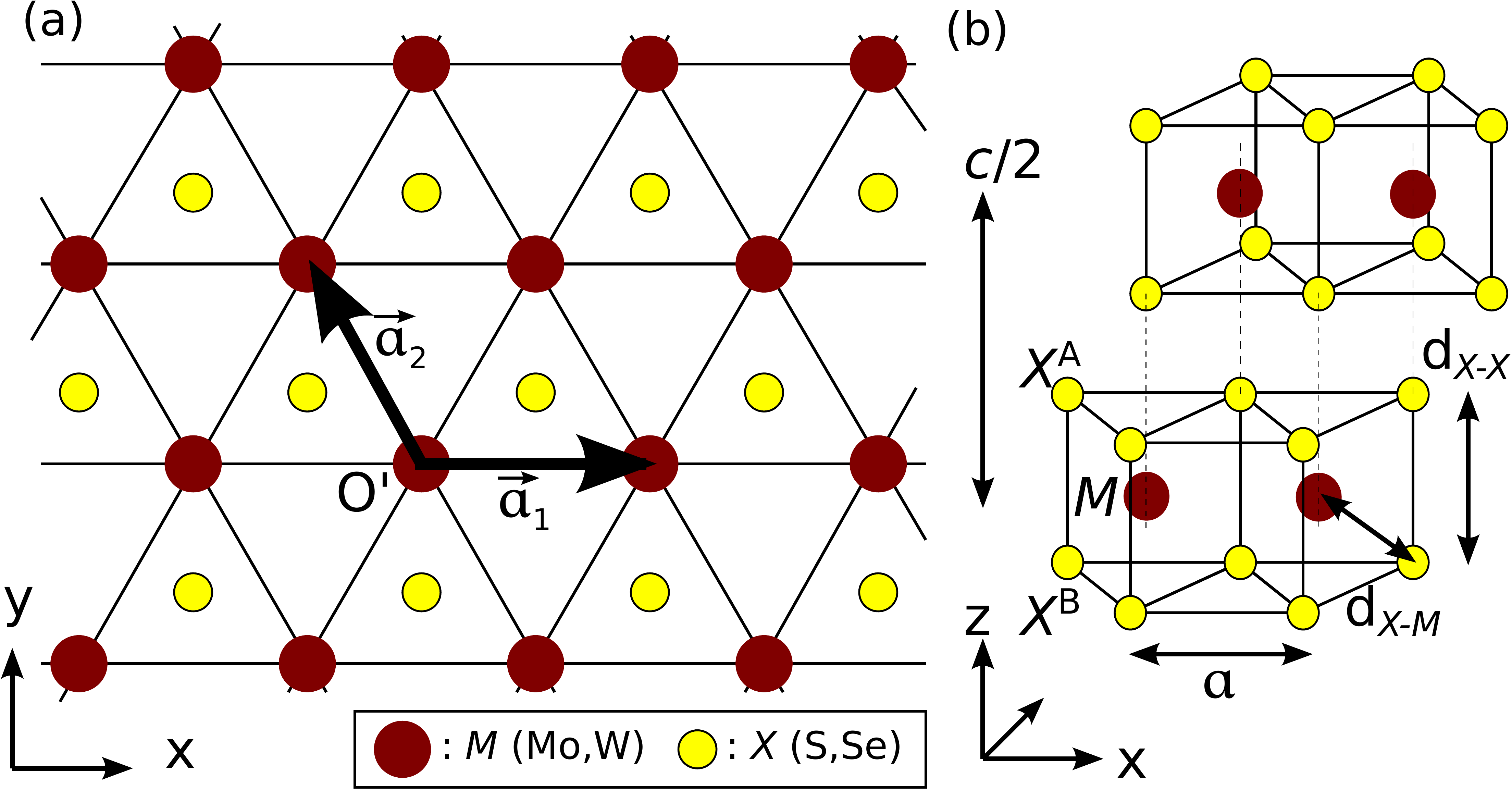}
\caption{(a) Top view of a single layer TMDC crystal structure with hexagonal primitive vectors $\vec{a}_1$ and $\vec{a}_2$. The smaller yellow circles represent the two chalcogen atoms separated by $d_{\rm X-X}$ in the z direction and the larger brown circles the metal atoms. (b) Perspective side view for the TMDC 2H crystal structure composed of X$^{\rm A}$-M-X$^{\rm B}$ monolayers with $a$ the in-plane lattice constant and $c/2$ the separation between two units along the $\hat{z}$-axis.}
\label{fig:TMDC_crystal}
\end{figure}

\section{\label{sec:level1}Crystal Structure And Symmetry}

TMDC crystals appear in different forms, labeled according to the stacking between successive layers as 1H, 1T and 1T', the latter having a 2 $\times$ 1 reconstruction of the planar unit cell\cite{TMDC_TI1}. The electronic properties are tied to the underlying crystal structure. The semiconducting 1H monolayer crystal structure is shown in Fig. \ref{fig:TMDC_crystal} and is the focus of this work. The chalcogen atoms, labeled as X$^{\rm A}$/X$^{\rm B}$ at the top/bottom, form a hexagonal lattice in each layer and project onto the same position in the plane of the middle layer of metal (M) atoms. As seen in Fig. \ref{fig:TMDC_crystal}(b), the M atoms have trigonal prismatic coordination. The monolayer unit is characterized by the hexagonal lattice constant $a$ and the projected distance from X to the middle layer d$_{\rm X-X}$/2, which are given in Table \ref{table:crystal}. In our convention, the primitive vectors are $\vec{a}_1=a\hat{x}$ and $\vec{a}_2=a(-\frac{1}{2}\hat{x}+\frac{\sqrt{3}}{2}\hat{y})$ as shown in Fig. \ref{fig:TMDC_crystal}, with reciprocal space vectors $\vec{b}_1=\frac{2\pi}{a}(\hat{x}+\frac{1}{\sqrt{3}}\hat{y})$ and $\vec{b}_2=\frac{4\pi}{a\sqrt{3}}\hat{y}$. The BZ has special \textit{k}-points, $\Gamma$ at the center, M$=\frac{1}{2}\vec{b}_1$, K$_\pm=\pm \frac{1}{3}(2\vec{b}_1-\vec{b}_2)$. 

The D$_{3h}$ symmetry group for the crystal contains a mirror symmetry in the xy plane ($\mathcal{M}_1$), a mirror symmetry in the yz plane ($\mathcal{M}_2$) centered at each atomic position and a three-fold rotation symmetry ($\mathcal{R}_3$); these symmetries are relevant to our hamiltonian construction. For multilayer TMDCs, there are various kinds of crystal structures, depending on the stacking along the \textit{c}-axis, with periodic vector $\vec{a}_3=c\hat{z}$ where $c$ is the distance between repeating monolayer units. In this work, we focus on the 2H stacking with two monolayer units in a unit cell with hexagonal symmetry, shown in Fig. \ref{fig:TMDC_crystal} (b), with the two units related by a mirror symmetry $\mathcal{M}_3$ in the xz plane, or equivalently by a $\pi$ rotation around the $\hat{z}$-axis. In a monolayer unit or a 2H stack with odd number of layers, the inversion symmetry is explicitly broken while the symmetry is restored in a 2H stack of even number of layers.

\begin{table}[ht!]
\caption{The in-plane lattice constant ($a$), unit cell size along the axis perpendicular to the plane ($c$), distance along the plane-normal direction between chalcogen layers (d$_\textrm{X-X}$) and nearest neighbor bond between metal and chalcogen atoms (d$_\textrm{X-M}$), all in \AA, for the TMDCs considered here. Numbers in brackets and $c$ are experimental bulk values\cite{TMDC_exp_crystal}. }
\label{table:crystal}
\centering
\vspace{1 mm}
\begin{tabular}{|c|c|c|c|c|}
\hline
\hline
                   & {\bf MoS$_2$}                    & {\bf MoSe$_2$}                   & {\bf WS$_2$}                     & {\bf WSe$_2$}                    \\ \hline
$a$(\AA)           & 3.18 {[}3.16{]}                  & 3.32 {[}3.29{]}                  & 3.18 {[}3.15{]}                  & 3.32 {[}3.28{]}                  \\ \hline
$c$(\AA)           & \multicolumn{1}{r|}{{[}12.29{]}} & \multicolumn{1}{r|}{{[}12.90{]}} & \multicolumn{1}{r|}{{[}12.32{]}} & \multicolumn{1}{r|}{{[}12.96{]}} \\ \hline
d$_{\rm X-X}$(\AA) & 3.13 {[}3.17{]}                  & 3.34 {[}3.33{]}                  & 3.14 {[}3.14{]}                  & 3.35 {[}3.34{]}                  \\ \hline
d$_{\rm X-M}$(\AA) & 2.41 {[}2.42{]}                  & 2.54 {[}2.52{]}                  & 2.42 {[}2.40{]}                  & 2.55 {[}2.53{]}                  \\ \hline
\hline
\end{tabular}
\end{table}

\section{\label{sec:level1}Monolayer Tight-Binding hamiltonian}

\subsection{\label{sec:level2}Truncated Tight-Binding Hamiltonian (TBH)}

In the FTBH, the number of neighbor couplings included depends on the size of reciprocal space sampling in the Wannier construction. However, the couplings are short-ranged and fall off exponentially with the neighbor distance. Starting from a fully converged FTBH, we can truncate the hamiltonian to retain only the first-neighbor terms and partial second-neighbor terms to improve the accuracy. This TBH captures the essential features of the bands and their orbital character. To construct such a hamiltonian properly, we need to identify the relevant atomic orbitals and the symmetries of the crystal. For TMDC materials, the relevant atomic orbital basis for a monolayer are the \textit{d}(\textit{p}) orbitals of M(X) atoms, labeled as:

\begin{equation}
\label{eqn:basis_chi}
\begin{split}
\hat{\psi}^{\dagger}_{pd}=&[\hat{d}^{\dagger}_{\rm z^2},\hat{d}^{\dagger}_{\rm xy},\hat{d}^{\dagger}_{\rm x^2-y^2},\hat{d}^{\dagger}_{\rm xz},\hat{d}^{\dagger}_{\rm yz}, \\
& \hat{p}^{\rm A \dagger}_{\rm x},\hat{p}^{\rm A \dagger}_{\rm y},\hat{p}^{\rm A \dagger}_{\rm z},\hat{p}^{\rm B \dagger}_{\rm x},\hat{p}^{\rm B \dagger}_{\rm y},\hat{p}^{\rm B \dagger}_{\rm z}]
\end{split}
\end{equation}


\begin{table}[ht!]
\caption{Odd and even basis sectors (superscripts (o) and (e)) and their symmetries under xy mirror reflection ($\mathcal{M}_1$) and yz mirror reflection ($\mathcal{M}_2$) in terms of atomic Wannier orbitals: the $d$-like orbitals of M atoms and the $\textit{p}^A$/$\textit{p}^B$ orbitals of X atoms located on the top/bottom layer.}
\label{table:basis_list_sym}
\centering
\vspace{1 mm}
\begin{tabular}{ |c|c|c|c|c| } 
\hline
Index & Basis Function & $\mathcal{M}_1$ & $\mathcal{M}_2$\\
\hline
1 & $d_{xz}^{\rm (o)}=d_{xz}$ & $-$ & $-$ \\ 
2 & $d_{yz}^{\rm (o)}=d_{yz}$ & $-$ & $+$ \\ 
3 & $p_z^{\rm (o)}=\frac{1}{\sqrt{2}}(p_z^{A}+p_z^{B})$ & $-$ & $+$  \\ 
4 & $p_x^{\rm (o)}=\frac{1}{\sqrt{2}}(p_x^{A}-p_x^{B})$ & $-$ & $-$ \\ 
5 & $p_y^{\rm (o)}=\frac{1}{\sqrt{2}}(p_y^{A}-p_y^{B})$ & $-$ & $+$ \\ 
\hline
6 & $d_{z^2}^{\rm (e)}=d_{z^2}$ & $+$ &  $+$ \\
7 & $d_{xy}^{\rm (e)}=d_{xy}$ & $+$ & $-$  \\
8 & $d_{x^2-y^2}^{\rm (e)}=d_{x^2-y^2}$ & $+$ & $+$\\
9 & $p_z^{\rm (e)}=\frac{1}{\sqrt{2}}(p_z^{A}-p_z^{B})$ & $+$ & $+$ \\
10 & $p_x^{\rm (e)}=\frac{1}{\sqrt{2}}(p_x^{A}+p_x^{B})$ & $+$ & $-$\\
11 & $p_y^{\rm (e)}=\frac{1}{\sqrt{2}}(p_y^{A}+p_y^{B})$ & $+$ & $+$ \\
\hline
\end{tabular}
\end{table}

The relevant symmetry operations of the monolayer include the mirror symmetries $\mathcal{M}_1$ and $\mathcal{M}_2$ and three-fold rotations $\mathcal{R}_3$. They are used to classify the matrix elements and reduce the number of independent parameters in the hamiltonian. First, the states are classified as odd or even under xy mirror symmetry $\mathcal{M}_1$\cite{mos2_tb_dft1}. The new basis that embodies this symmetry is:

\begin{equation}
\label{eqn:basis_phi}
\begin{split}
\hat{\phi}^{\dagger}_{\rm eo} &= [\hat{d}_{\rm xz}^{\rm (o) \dagger}, \hat{d}_{\rm yz}^{\rm (o) \dagger}, \hat{p}_z^{\rm (o) \dagger},\hat{p}_x^{\rm (o) \dagger}, \hat{p}_y^{\rm (o) \dagger}, \\
&\hat{d}_{\rm z^2}^{\rm (e) \dagger},\hat{d}_{\rm xy}^{\rm (e) \dagger},\hat{d}_{\rm x^2-y^2}^{\rm (e) \dagger}, \hat{p}_z^{\rm (e) \dagger}, \hat{p}_x^{\rm (e) \dagger}, \hat{p}_y^{\rm (e) \dagger}]
\end{split}
\end{equation} and contains five states in the odd sector and six in the even sector (see Table \ref{table:basis_list_sym} for details). Under $\mathcal{M}_1$ symmetry the pair of X atoms in a unit cell is treated as a single composite atom with odd or even states. Because the hamiltonian commutes with $\mathcal{M}_1$, it can be written in even and odd diagonal block form and there are no mixing terms between the two blocks which break the $\mathcal{M}_1$ symmetry. The two basis sets are linked by the unitary transformation $\mathcal{T}$, where $\hat{\phi}_{\rm eo}^{\dagger}=\mathcal{T}\hat{\psi}_{pd}^{\dagger}$. When there are multiple layers with interaction between layers, the transformation between the two sets of basis is needed to determine the inter-layer hopping terms. If spin is included, the generalized even/odd sectors of the Hilbert space can be defined (see the discussion on spin-orbit coupling). The minimal spinless hamiltonian for a monolayer is given by:

\begin{equation}
\hat{H}^{\rm (1L)}=\sum_{i,j,\vec{k}} \hat{\phi}^{\dagger}_i(\vec{k}) H_{i,j}^{\rm (1L)}(\vec{k}) \hat{\phi}_j (\vec{k})
\end{equation}


\begin{figure}
\centering
\includegraphics[width=0.5\textwidth]{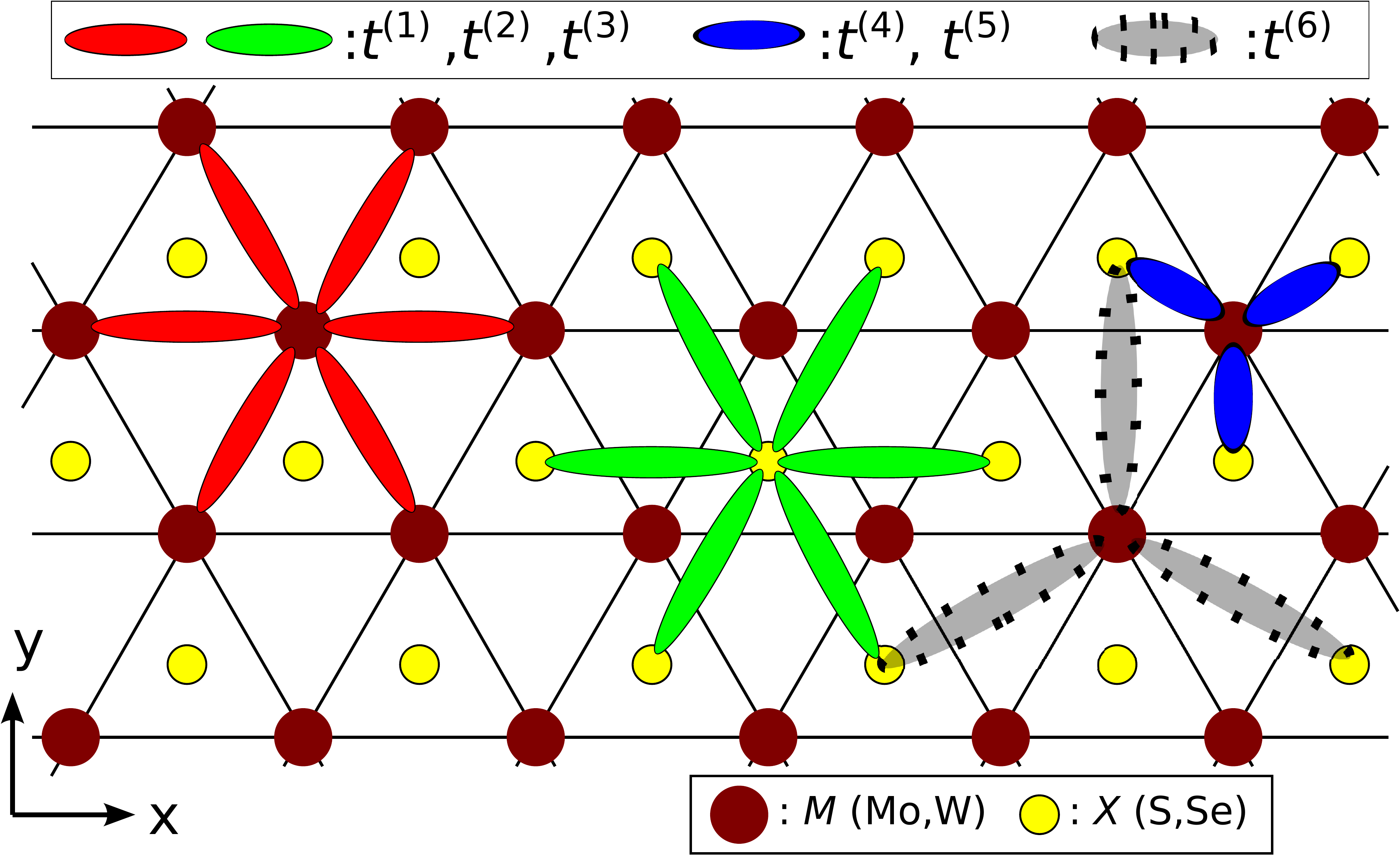}
\caption{Schematic diagram for all hopping terms in the TBH: M-M coupling (red) with six first-neighbor pairs; X-X coupling (green) with six first-neighbor pairs; X-M coupling (blue) with three first-neighbor pairs. Additional couplings between three second-neighbor X-M pairs (grey dashed) are used to improve the accuracy. The top bar indicates the labels of parameters that correspond to the different types of hopping terms, $t^{(n)}$, $n=1,\ldots,6$. }
\label{fig:TMDC_miniTB}
\end{figure}


This minimal tight-binding hamiltonian retains only the diagonal energy terms and the hopping terms as shown in Fig. \ref{fig:TMDC_miniTB}: M-M hoppings (red) stand for the coupling from the odd (even) states on one M atom at to the odd (even) states on its six first-neighbor M atoms and X-X hoppings (green) are defined similarly for the six first-neighbor pairs of X atoms. X-M hoppings (blue) link states on M atoms to states at the three first-neighbor X atoms. The M-X hoppings are Hermitian conjugates of X-M hoppings. The grey dotted hoppings are additional X-M second neighbor terms that can be included to improve the accuracy for low-energy bands. All the relevant vectors that connect orbitals in the different hopping terms of the hamiltonian are defined in Table \ref{table:hop_vectors}.

\begin{table}[ht!]
\caption{The hopping vectors in the TBH}
\label{table:hop_vectors}
\centering
\vspace{1 mm}
\begin{tabular}{ |c|c|c|c|c| } 
\hline
$t^{(1)},t^{(2)},t^{(3)}$ & $\vec{\delta}_1=\vec{a}_1$, $\vec{\delta}_2=\vec{a}_1+\vec{a}_2$, $\vec{\delta}_3=\vec{a}_2$ \\
\hline
$t^{(4)},t^{(5)}$ & $\vec{\delta}_4=-(2\vec{a}_1+\vec{a}_2)/3$, $\vec{\delta}_5=(\vec{a}_1+2\vec{a}_2)/3$ \\
 & $\vec{\delta}_6=(\vec{a}_1-\vec{a}_2)/3$ \\
\hline
$t^{(6)}$ & $\vec{\delta}_{7}=-2(\vec{a}_1+2\vec{a}_2)/3$, $\vec{\delta}_{8}=2(2\vec{a}_1+\vec{a}_2)/3$ \\
& $\vec{\delta}_{9}=2(\vec{a}_2-\vec{a}_1)/3$ \\
\hline
\end{tabular}
\end{table}

We present next the matrix elements of the hamiltonian after Fourier transformation to the \textit{k} space in the BZ. We use the notation $t_{i,j}^{(s)}=<\phi_i|\hat{H}| \phi_j >$ for the hopping matrix element from the state $\phi_j$ to the state $\phi_i$ with the label \textit{s} specifying the type and the distance between the two atoms where $\phi_i$ and $\phi_j$ reside. The diagonal part for the hamiltonian contains $\epsilon_i$, the on-site energy of orbitals, and the hopping terms between orbitals of the same type at first-neighbor positions. The diagonal terms of the tight-binding hamiltonian take the form:

\begin{equation}
\label{eqn:ham1}
\begin{split}
H_{i,i}^{\rm (1L)}(\vec{k})&=\epsilon_i+2t_{i,i}^{(1)}\cos{\vec{k} \cdotp \vec{\delta}_1} \\
& +2t_{i,i}^{(2)}[\cos(\vec{k} \cdotp \vec{\delta}_2)+\cos(\vec{k} \cdotp \vec{\delta}_3)]
\end{split}
\end{equation} The off-diagonal hopping matrix elements between the same type of atoms (M-M and X-X) can be classified into two categories depending on the symmetry of the \textit{i} and \textit{j} orbitals under $\mathcal{M}_2$: for (\textit{i},\textit{j})=(3,5),(6,8),(9,11), the symmetry is ($+$), giving
\begin{equation}
\label{eqn:ham2}
\begin{split}
H_{i,j}^{\rm (1L)}(\vec{k})=&2t_{i,j}^{(1)}\cos{\vec{k} \cdotp \vec{\delta}_1}+t_{i,j}^{(2)}[e^{-i\vec{k} \cdotp \vec{\delta}_2}+e^{-i\vec{k}\cdotp \vec{\delta}_3}] \\
&+t_{i,j}^{(3)}[e^{i\vec{k}\cdotp \vec{\delta}_2}+e^{i \vec{k} \cdotp \vec{\delta}_3}]
\end{split}
\end{equation} while for (\textit{i},\textit{j})=(1,2),(3,4),(4,5),(6,7),(7,8),(9,10),(10,11), the symmetry is ($-$), giving
\begin{equation}
\label{eqn:ham3}
\begin{split}
H_{i,j}^{\rm (1L)}(\vec{k})=&-2it_{i,j}^{(1)}\sin{\vec{k} \cdotp \vec{\delta}_1}+t_{i,j}^{(2)}[e^{-i\vec{k}\cdotp \vec{\delta}_2}-e^{-i\vec{k}\cdotp \vec{\delta}_3}] \\
&+t_{i,j}^{(3)}[-e^{i\vec{k}\cdotp \vec{\delta}_2}+e^{i\vec{k}\cdotp \vec{\delta}_3}]
\end{split}
\end{equation}

A different type of hopping connects M and X atoms. Each M atom has three first neighbor X pairs. The matrix elements can still be classified by the symmetry of orbitals \textit{i} and \textit{j} under $\mathcal{M}_2$, that is, for the pairs (\textit{i},\textit{j})=(3,1), (5,1), (4,2), (10,6), (9,7), (11,7), (10,8), the symmetry is ($+$), giving 
\begin{equation}
\label{eqn:ham4}
H_{i,j}^{\rm (1L)}(\vec{k})=t_{i,j}^{(4)}[e^{i\vec{k}\cdotp \vec{\delta}_4}-e^{i\vec{k}\cdotp \vec{\delta}_6}]
\end{equation} while for the pairs (\textit{i},\textit{j})=(4,1), (3,2), (5,2), (9,6), (11,6), (10,7), (9,8), (11,8), the symmetry is ($-$), giving
\begin{equation}
\label{eqn:ham5}
H_{i,j}^{\rm (1L)}(\vec{k})=t_{i,j}^{(4)}[e^{i\vec{k}\cdotp \vec{\delta}_4}+e^{i\vec{k}\cdotp \vec{\delta}_6}]+t_{i,j}^{(5)}e^{i\vec{k}\cdotp \vec{\delta}_5}
\end{equation} The hermitian character of the hamiltonian requires $H_{i,j}^{\rm (1L)}(\vec{k})=H^{\rm (1L)}_{j,i}(\vec{k})^*$ and otherwise unassigned $H^{\rm (1L)}_{i,j}(\vec{k})$ terms are zero.

The above minimal tight-binding hamiltonian has 86 parameters for all $\epsilon_i$, $t_{i,j}^{(1)}$, $t_{i,j}^{(2)}$, $t_{i,j}^{(3)}$, $t_{i,j}^{(4)}$ and $t_{i,j}^{(5)}$ under the $\mathcal{M}_1$ and $\mathcal{M}_2$ symmetry classifications. The three-fold symmetry $\mathcal{R}_3$ can be used to further reduce the size of the parameter space by identifying equivalent coupling directions: with this simplification, only a subset of $\epsilon_i$ and all $t_{i,j}^{(1)}$, $t_{i,j}^{(5)}$ are independent parameters while $t_{i,j}^{(2)}$, $t_{i,j}^{(3)}$ and $t_{i,j}^{(4)}$ can be expressed as linear combinations of those,  which reduces the number of free parameters to 36. The relationships that express these symmetry-imposed simplifications are given in Appendix A.

\begin{figure}
\centering
\includegraphics[width=0.48\textwidth]{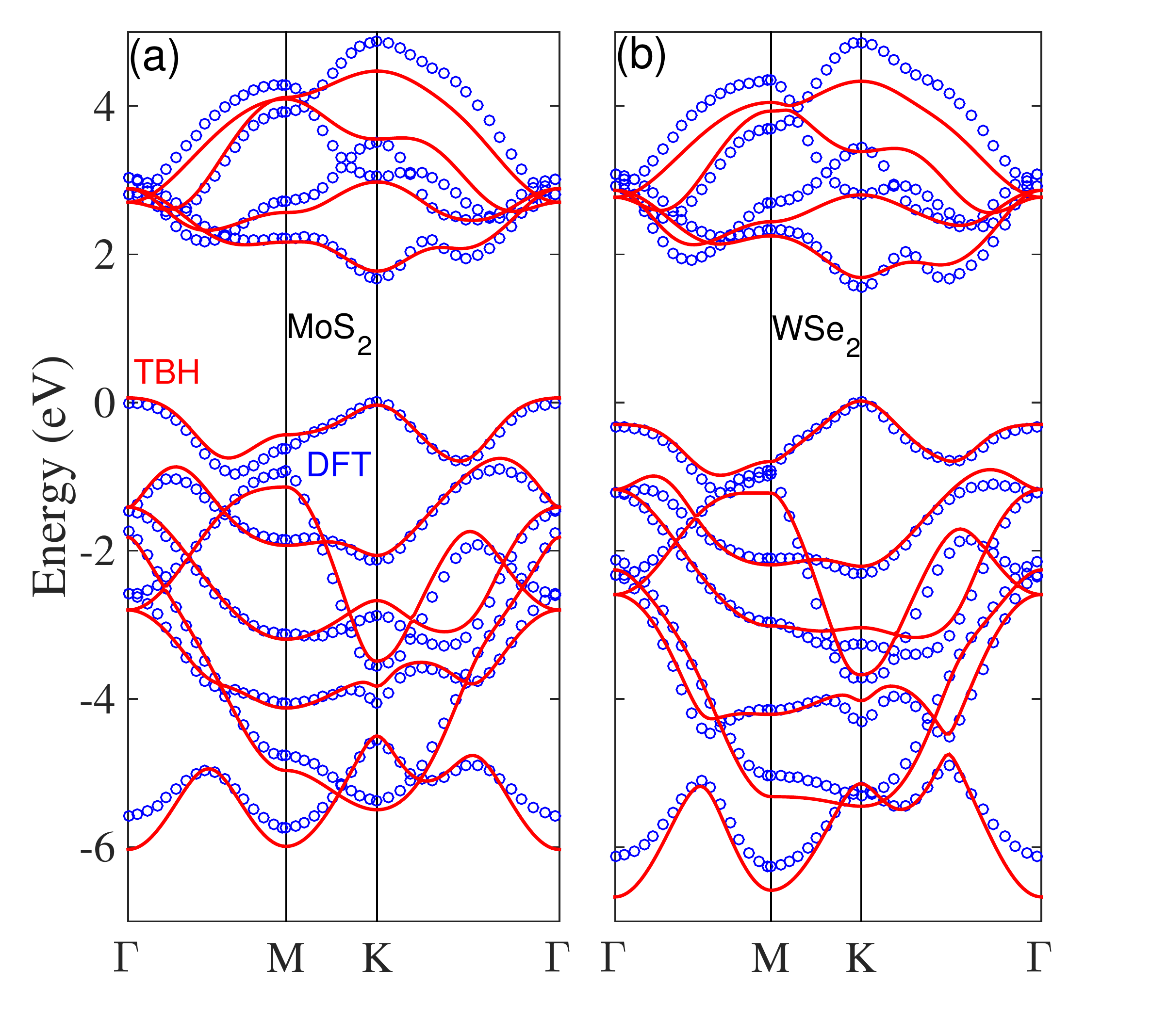}
\caption{TBH (red lines) band structure along $\Gamma$-M-K-$\Gamma$ and the comparison with DFT results (blue circles) for monolayer spinless (a) MoS$_2$ (b) WSe$_2$.}
\label{fig:TMDC_mono}
\end{figure}

In most applications, the only relevant degrees of freedom involved in physical processes are the lowest conduction band and highest valence band at K. The orbital character of these bands are given in Table \ref{table:orbital_sym}. They are even states under $\mathcal{M}_1$. 

The bands in the above minimal tight-binding hamiltonian can be improved in an efficient way by including terms beyond first-neighbor couplings. Based on an analysis of the orbital character of low-energy bands (see next section), our choice is to add the dominant second-neighbor X-M coupling terms between the even states (described by the grey dashed lines in Fig. \ref{fig:TMDC_miniTB}). There are four additional independent parameters in $\hat{H}^{' \rm (1L)}$ needed to include these coupling terms, labeled $t_{9,6}^{(6)}$, $t_{11,6}^{(6)}$, $t_{9,8}^{(6)}$, $t_{11,8}^{(6)}$, and the correction to the TBH reads:
\begin{equation}
\hat{H}^{'\rm (1L)}=\sum_{i,j,\vec{k}} \hat{\phi}^{\dagger}_i(\vec{k}) H_{i,j}^{'\rm (1L)}(\vec{k}) \hat{\phi}_j (\vec{k})
\end{equation} with the hermitian matrix elements:

\begin{equation}
\label{eqn:ham6}
\begin{split}
H^{' \rm (1L)}_{9,6}(\vec{k})&= t_{9,6}^{(6)}(e^{i\vec{k} \cdotp \vec{\delta}_{7}}+e^{i\vec{k} \cdotp \vec{\delta}_{8}}+e^{i\vec{k} \cdotp \vec{\delta}_{9}}) \\
H^{' \rm (1L)}_{11,6}(\vec{k})&= t_{11,6}^{(6)}(e^{i\vec{k} \cdotp \vec{\delta}_{7}}-\frac{1}{2}e^{i\vec{k} \cdotp \vec{\delta}_{8}}-\frac{1}{2}e^{i\vec{k} \cdotp \vec{\delta}_{9}}) \\
H^{' \rm (1L)}_{10,6}(\vec{k})&=\frac{\sqrt{3}}{2} t_{11,6}^{(6)}(-e^{i\vec{k} \cdotp \vec{\delta}_{8}}+e^{i\vec{k} \cdotp \vec{\delta}_{9}})  \\
H^{' \rm (1L)}_{9,8}(\vec{k})&= t_{9,8}^{(6)}(e^{i\vec{k} \cdotp \vec{\delta}_{7}}-\frac{1}{2}e^{i\vec{k} \cdotp \vec{\delta}_{8}}-\frac{1}{2}e^{i\vec{k} \cdotp \vec{\delta}_{9}})  \\
H^{' \rm (1L)}_{9,7}(\vec{k})&=\frac{\sqrt{3}}{2} t_{9,8}^{(6)}(-e^{i\vec{k} \cdotp \vec{\delta}_{8}}+e^{i\vec{k} \cdotp \vec{\delta}_{9}}) \\
H^{' \rm (1L)}_{10,7}(\vec{k})&=\frac{3}{4} t_{11,8}^{(6)}(e^{i\vec{k} \cdotp \vec{\delta}_{8}}+e^{i\vec{k} \cdotp \vec{\delta}_{9}})\\
H^{' \rm (1L)}_{11,7}(\vec{k})&=H^{' \rm (1L)}_{10,8}(\vec{k})=\frac{\sqrt{3}}{4} t_{11,8}^{(6)}(e^{i\vec{k} \cdotp \vec{\delta}_{8}}-e^{i\vec{k} \cdotp \vec{\delta}_{9}}) \\
H^{' \rm (1L)}_{11,8}(\vec{k})&= t_{11,8}^{(6)}(e^{i\vec{k} \cdotp \vec{\delta}_{7}}+\frac{1}{4}e^{i\vec{k} \cdotp \vec{\delta}_{8}}+\frac{1}{4}e^{i\vec{k} \cdotp \vec{\delta}_{9}})
\end{split}
\end{equation}

The final TBH is defined as the sum of the two terms, $\hat{H}^{(\rm 1L,TBH)}=\hat{H}^{\rm (1L)}+\hat{H}^{' \rm (1L)}$. In Fig. \ref{fig:TMDC_mono}, we show the resulting band structure from the TBH for MoS$_2$ and WSe$_2$ in comparison with DFT results. We also note that even when the hamiltonian is truncated to retain only the first-neighbor terms (only $\hat{H}^{\rm (1L)}$), this minimal hamiltonian still preserves the essential features and orbital character of the bands. The tight-binding results for the valence bands are better than for the conduction bands. This is expected due to the proximity of other higher energy bands near the conduction bands to which we have restricted the model presented here. If the bands belong to the same symmetry group representation, the level anticrossing between these bands introduces orbital character mixing. This mixing generates higher order corrections in the effective hamiltonian when the irrelevant bands are integrated out. To account for these effects, higher order coupling terms which involve coupling between farther atoms are needed to obtain a more accurate description. 

To improve the underestimated band-gap value at the DFT level, we performed GW calculations and corrected the tight-binding hamiltonian parameters accordingly for MoS$_2$. The details and the scaling factors of the tight-binding parameters for the GW quasi-particle energies are given in Appendix B. For the rest of the discussion, we will focus on the TBH bands based on DFT results, which is adequate to bring out the physics.

\subsection{\label{sec:level2}Orbital Characters of Bands}

It is interesting to investigate the spinless orbital character of the bands at the high-symmetry $\Gamma$ and K$_\pm$ points, which are relevant for low-energy dynamics. These $k$ points have $\mathcal{R}_3$ symmetry, and the bands can be characterized by the symmetry group representations. To facilitate the classification of the states, we assign as labels the eigenvalue $m_z$ of the angular momentum operator $\hat{L}_z$: $d_{\pm 2}=\frac{1}{\sqrt{2}}(d_{x^2-y^2}\pm id_{xy})$, $d_{\pm 1}=\mp \frac{1}{\sqrt{2}}(d_{xz}\pm id_{yz})$, $d_{0}=d_{z^2}$, $p_{\pm 1}=\mp \frac{1}{\sqrt{2}}(p_x \pm ip_y)$, $p_{0}=p_z$. Under clockwise $\mathcal{R}_3$ symmetry with the axis of rotation centered at the atomic positions, these orbitals are invariant and have eigenvalues $e^{+i2\pi m_z /3}$. The orbital character of the bands at these special \textit{k} points is given in Table \ref{table:orbital_sym}, along with the coefficients $c_i$ ($\tilde{c}_i=\sqrt{1-c_i^2}$) for the four TMDC materials considered here.

\begin{widetext}

\centering

\begin{table}[h]
\centering
\caption{Orbital composition and numerical coefficients for the eleven bands at $\Gamma$, K$_+$ (states at K$_-$ are related to the states at K$_+$ by the time-reversal symmetry). In the subscript notation $\bar{m}$ stands for $-m$ ($m=$1,2). For the coefficients, we define $\tilde{c}_i=\sqrt{1-c_i^2}$. Bands 1-7 (8-11) are valence (conduction) bands.}
\label{table:orbital_sym}
\begin{tabular}{|c|c|lcccl|c|l|c|c|c|}
\cline{1-2} \cline{8-8} \cline{10-12}
{\bf Band} & {\bf $\Gamma$}                                              &                       & \multicolumn{1}{l}{}             & \multicolumn{1}{l}{}               & \multicolumn{1}{l}{}                & $\; \; \; $ & {\bf K$_+$}                                                &  &             & {\bf MoS$_2$} & {\bf MoSe$_2$} \\ \cline{1-2} \cline{8-8} \cline{10-12} 
{\bf 11}   & $i\tilde{c}_2d_{2}^{\rm (e)}-c_2p_{\bar{1}}^{\rm (e)}$      &                       & \multicolumn{1}{l}{}             & \multicolumn{1}{l}{}               & \multicolumn{1}{l}{}                &             & $i\tilde{c}_{8}d_{1}^{\rm (o)}-c_{8}p_0^{\rm (o)}$         &  & {\bf $c_8$} & 0.5263        & 0.5772         \\ \cline{1-2} \cline{8-8} \cline{10-12} 
{\bf 10}   & $i\tilde{c}_2d_{\bar{2}}^{\rm (e)}-c_2p_{1}^{\rm (e)}$      &                       & \multicolumn{1}{l}{}             & \multicolumn{1}{l}{}               & \multicolumn{1}{l}{}                &             & $\tilde{c}_{4}d_{\bar{1}}^{\rm (o)}+c_{4}p_{1}^{\rm (o)}$  &  & {\bf $c_7$} & 0.4255        & 0.4344         \\ \cline{1-2} \cline{8-8} \cline{10-12} 
{\bf 9}    & $\tilde{c}_3d_{1}^{\rm (o)}+c_3p_{1}^{\rm (o)}$             &                       & \multicolumn{1}{l}{}             & \multicolumn{1}{l}{}               & \multicolumn{1}{l}{}                &             & $\tilde{c}_{7}d_{\bar{2}}^{\rm (e)}-c_{7}p_0^{\rm (e)}$    &  & {\bf $c_6$} & 0.4432        & 0.4204         \\ \cline{1-2} \cline{4-6} \cline{8-8} \cline{10-12} 
{\bf 8}    & $\tilde{c}_3d_{\bar{1}}^{\rm (o)}+c_3p_{\bar{1}}^{\rm (o)}$ & \multicolumn{1}{l|}{} & \multicolumn{1}{c|}{}            & \multicolumn{1}{c|}{{\bf MoS$_2$}} & \multicolumn{1}{c|}{{\bf MoSe$_2$}} &             & $i\tilde{c}_{5}d_{0}^{\rm (e)}+c_{5}p_{\bar{1}}^{\rm (e)}$ &  & {\bf $c_5$} & 0.4026        & 0.4012         \\ \cline{1-2} \cline{4-6} \cline{8-8} \cline{10-12} 
{\bf 7}    & $\tilde{c}_1 d_{0}^{\rm (e)}-c_1 p_0^{\rm (e)}$             & \multicolumn{1}{l|}{} & \multicolumn{1}{c|}{{\bf $c_3$}} & \multicolumn{1}{c|}{0.7239}        & \multicolumn{1}{c|}{0.7779}         &             & $i\tilde{c}_{6}d_{2}^{\rm (e)}+c_{6}p_{1}^{\rm (e)}$       &  & {\bf $c_4$} & 0.6268        & 0.6149         \\ \cline{1-2} \cline{4-6} \cline{8-8} \cline{10-12} 
{\bf 6}    & $c_3d_{1}^{\rm (o)}-\tilde{c}_3p_{1}^{\rm (o)}$             & \multicolumn{1}{l|}{} & \multicolumn{1}{c|}{{\bf $c_2$}} & \multicolumn{1}{c|}{0.8010}        & \multicolumn{1}{c|}{0.8234}         &             & $c_{8}d_{1}^{\rm (o)}-i\tilde{c}_{8}p_0^{\rm (o)}$         &  &             & {\bf WS$_2$}  & {\bf WSe$_2$}  \\ \cline{1-2} \cline{4-6} \cline{8-8} \cline{10-12} 
{\bf 5}    & $c_3d_{\bar{1}}^{\rm (o)}-\tilde{c}_3p_{\bar{1}}^{\rm (o)}$ & \multicolumn{1}{l|}{} & \multicolumn{1}{c|}{{\bf $c_1$}} & \multicolumn{1}{c|}{0.5711}        & \multicolumn{1}{c|}{0.5348}         &             & $p_{\bar{1}}^{\rm (o)}$                                    &  & {\bf $c_8$} & 0.4826        & 0.5394         \\ \cline{1-2} \cline{4-6} \cline{8-8} \cline{10-12} 
{\bf 4}    & $p_0^{\rm (o)}$                                             & \multicolumn{1}{l|}{} & \multicolumn{1}{c|}{}            & \multicolumn{1}{c|}{{\bf WS$_2$}}  & \multicolumn{1}{c|}{{\bf WSe$_2$}}  &             & $c_{7}d_{\bar{2}}^{\rm (e)}+\tilde{c}_{7}p_0^{\rm (e)}$    &  & {\bf $c_7$} & 0.3883        & 0.3988         \\ \cline{1-2} \cline{4-6} \cline{8-8} \cline{10-12} 
{\bf 3}    & $c_2d_{2}^{\rm (e)}-i\tilde{c}_2p_{\bar{1}}^{\rm (e)}$      & \multicolumn{1}{l|}{} & \multicolumn{1}{c|}{{\bf $c_3$}} & \multicolumn{1}{c|}{0.6698}        & \multicolumn{1}{c|}{0.7322}         &             & $c_{6}d_{2}^{\rm (e)}+i\tilde{c}_{6}p_{1}^{\rm (e)}$       &  & {\bf $c_6$} & 0.4643        & 0.4450         \\ \cline{1-2} \cline{4-6} \cline{8-8} \cline{10-12} 
{\bf 2}    & $c_2d_{\bar{2}}^{\rm (e)}-i\tilde{c}_2p_{1}^{\rm (e)}$      & \multicolumn{1}{l|}{} & \multicolumn{1}{c|}{{\bf $c_2$}} & \multicolumn{1}{c|}{0.7850}        & \multicolumn{1}{c|}{0.8077}         & $\; \; \; $ & $c_{5}d_{0}^{\rm (e)}+i\tilde{c}_{5}p_{\bar{1}}^{\rm (e)}$ &  & {\bf $c_5$} & 0.3564        & 0.3536         \\ \cline{1-2} \cline{4-6} \cline{8-8} \cline{10-12} 
{\bf 1}    & $c_1d_{0}^{\rm (e)}+\tilde{c}_1p_0^{\rm (e)}$               & \multicolumn{1}{l|}{} & \multicolumn{1}{c|}{{\bf $c_1$}} & \multicolumn{1}{c|}{0.5729}        & \multicolumn{1}{c|}{0.5414}         & $\; \; \; $ & $c_{4}d_{\bar{1}}^{\rm (o)}-\tilde{c}_{4}p_{1}^{\rm (o)}$  &  & {\bf $c_4$} & 0.6291        & 0.6173         \\ \cline{1-2} \cline{4-6} \cline{8-8} \cline{10-12} 
\end{tabular}
\end{table}

\end{widetext}

At $\Gamma$, there are four pairs of doubly degenerate states which form the two-dimensional representation under the symmetry group of the crystal. Each of the states are chosen to be invariant under $\mathcal{R}_3$ and the symmetry demands $m_p-m_d=3n$ for the state mixing, where $m$ is the eigenvalue for the orbital angular momentum $\hat{L}_z$ and \textit{n} is an integer. Under yz mirror symmetry, $m$ goes to $-m$, and this results in the two dimensional degenerate subspace at $\Gamma$.

At K$_\pm$, all states are non-degenerate. Because the wavevector $\vec{k}$ at K$_\pm$ is not zero, we have to consider the symmetry transformations of the orbitals and of the spatial wavefunction. We take the rotational axis of $\mathcal{R}_3$ to be located at the position of M atoms. Since X atoms are not at the center, they will acquire an additional phase factor, $e^{\pm i2\pi / 3}$, under this symmetry. When both phase factors from orbital and spatial transformations are included, we arrive at the constraint $m_p \pm 1 -m_d =3n'$ for each state and the eigenvalues of $\mathcal{R}_3$ can be determined. States at K$_-$ are related to states at K$_+$ by complex conjugation of the wavefunctions in the $\hat{\psi}$ basis, dictated by time-reversal symmetry for spinless particles. 

As an example of the implications of these symmetry considerations, we mention that the highest valence band and lowest conduction band at K$_\pm$ are coupled by photons of polarization $\sigma_\pm$ due to the broken inversion symmetry and the chiral selection rule (see the application to optical transitions in Sec. VII). This allows the optical control of valley degrees of freedom and the breaking of valley degeneracy by circularly polarized light\cite{valley_break_opt}. The $\pi$ light with polarization perpendicular to the layer, which has odd symmetry under $\mathcal{M}_1$, cannot couple these two even states\cite{pi_polarization}. In forming stacks of TMDC layers, the bands with strong $p_0$ character should be affected most by the interlayer hybridization. This implies that the highest valence band at $\Gamma$ is sensitive to the interlayer coupling. On the other hand, the low-energy bands at K$_\pm$ are not sensitive to interlayer couplings because they are dominated by the $d$ orbitals of M atoms, which are deeper inside the monolayer unit.

\subsection{\label{sec:level2}Spin-Orbit coupling}
Up to this point, we have not included the spin degrees of freedom. Due to the lack of inversion symmetry in TMDCs, spin-splitting of bands generally occurs over the entire BZ. However, the bands are still doubly degenerate at the time-reversal invariant points $\Gamma$ and M due to the Kramers' theorem and at special $k$ points, such as along the $\Gamma$-M direction, which are protected by crystal mirror symmetry. In TMDCs, the low-energy bands at K$_\pm$ are not spin-degenerate and hence the spin is crucial for understanding the low-energy dynamics. The even/odd sectors under $\mathcal{M}_1$ mirror symmetry must be redefined to incorporate the spin degrees of freedom. Under $\mathcal{M}_1$ xy mirror operation, $\vec{\hat{S}}$ will flip the signs of $\hat{S}_x$ and $\hat{S}_y$ components but not of $\hat{S}_z$. Hence, $\mathcal{M}_1$ mirror symmetry acts as a $\pi$ rotation along the $\hat{z}$-axis, $\hat{\mathcal{O}}_{\mathcal{M}_1}=e^{i\pi\hat{S}_z/\hbar}$. The spin up (down) state is an eigenstates with eigenvalue\cite{bi_sb_surface} $+i$ ($-i$). The generalized $\mathcal{M}_1$ operation is the product of its action on the orbital and on the spin degrees of freedom. In this extended Hilbert space, the states of even orbitals with spin up (down) and the states of odd orbitals with spin down (up) are in the same enlarged $+i$ ($-i$) sector under $\mathcal{M}_1$. These generalized $\pm i$ sectors dictate which set of states can be mixed in the hamiltonian and there are no mixing terms between the generalized $+i$ and $-i$ sectors.

To incorporate this spin-orbit coupling effect in the tight-binding hamiltonian, we approximate the contribution for spin-orbit interaction by the atomic term $\lambda_{SO}^{\rm M/X} \vec{L} \cdotp \vec{S}$ for each individual M and X atom\cite{tmdc_socls}. In this expression, $\vec{L}=\hat{L}_x, \hat{L}_y, \hat{L}_z$($\vec{S}=\hat{S}_x, \hat{S}_y, \hat{S}_z$) stands for the orbital (spin) angular momentum operators, with $\lambda_{SO}$ the strength for the spin-orbit coupling term which depends on the atomic species (see Table \ref{table:SOC}). We obtain the values for various M and X atoms from the energy splitting of a single atom in a large unit cell as calculated by DFT when spin-orbit coupling is included in the Kohn-Sham equations. The Hilbert space for the monolayer hamiltonian with spin-orbit coupling (SOC) is the direct product space of Wannier orbitals and spin degrees of freedom, with 22 states in total. The TBH + SOC hamiltonian is given by:

\begin{figure}
\centering
\includegraphics[width=0.5\textwidth]{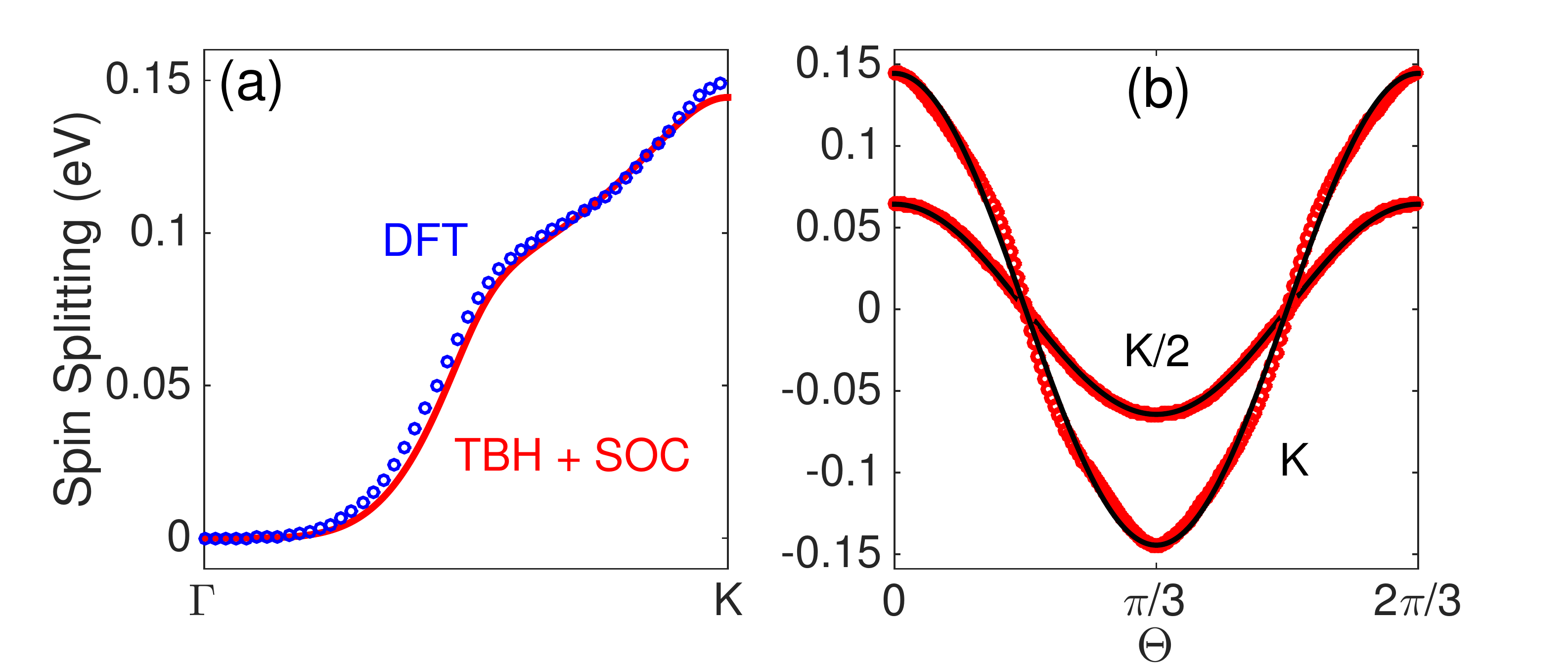}
\caption{Spin splitting for the highest pair of valence bands in MoS$_2$: (a) along the $\Gamma$-K direction as obtained from TBH + SOC (red lines) and DFT (blue dots) calculations (b) The angular dependence with fixed $k_\parallel$=K/2 and $k_\parallel$=K of TBH + SOC (red dots); the black lines are the fitted $\cos(3\theta)$ functions.}
\label{fig:tbh_soc}
\end{figure}


\begin{equation}
\begin{split}
\label{eqn:ham_so}
&\hat{H}^{\rm (1L)}_{\rm SO}=\sum_{\vec{k}} [\hat{\phi}^{\dagger}_{\uparrow}(\vec{k}) H^{\rm (1L,TBH)}_{\uparrow\uparrow}(\vec{k}) \hat{\phi}_{\uparrow} (\vec{k})  \\
&+\hat{\phi}^{\dagger}_{\downarrow}(k) H^{\rm (1L,TBH)}_{\downarrow\downarrow}(\vec{k}) \hat{\phi}_{\downarrow} (\vec{k})+ \hat{\phi}^{\dagger}(\vec{k}) H_{\rm LS} \hat{\phi}(\vec{k})]
\end{split}
\end{equation} 
The diagonal blocks in the first term $H^{\rm (1L,TBH)}_{\uparrow \uparrow}=H^{\rm (1L,TBH)}_{\downarrow \downarrow}=H^{\rm (1L,TBH)}$ are the TBH described above. These are the spin-independent hopping processes. The effect of spin-orbit coupling, $H_{\rm LS}$, is incorporated by the on-site $\lambda_{SO} \vec{L} \cdotp \vec{S}$ term for each atom. Because it is an on-site term, it does not carry momentum dependence and is a constant matrix with the matrix elements:

\begin{equation}
\begin{split}
&<\phi_{i,\sigma}|H_{\rm LS}|\phi_{j,\sigma'}>  \\
&=<\phi_{i,\sigma}| (\lambda_{SO}^{\rm M} \vec{L}_{\rm M} +\lambda_{SO}^{\rm X} \vec{L}_{\rm X}^A+\lambda_{SO}^{\rm X} \vec{L}_{\rm X}^B)\cdotp \vec{S} |\phi_{j,\sigma'}>
\end{split}
\end{equation} It is straightforward to evaluate these matrix elements with the help of Clebsch-Gordan coefficients. The LS terms can also be written as $(\vec{L}_+ \vec{S}_- +\vec{L}_- \vec{S}_+)/2 +\vec{L}_z \vec{S}_z$, and the change in the evenness/oddness of the orbitals is always accompanied by the change of the spin state. This observation affirms the above classification of the generalized $+i$ and $-i$ sectors under $\mathcal{M}_1$.

To compare the TBH + SOC results with the full DFT results, we compute the spin splitting energy, $\Delta_k^v=\epsilon_{\uparrow}^v(k)-\epsilon_{\downarrow}^v(k)$, for the two highest valence bands which are a spin up and down pair. The spins remain doubly degenerate along the $\Gamma$-M direction and the largest splitting is along the $\Gamma$-K direction. In Fig. \ref{fig:tbh_soc} (a) the TBH + SOC spin splitting is plotted along $\Gamma$-K (in red line) and compared with the DFT results (in blue dots). The change in the spin splitting is also accompanied by the change in the underlying orbital composition from $\Gamma$ to K\cite{tmdc_soc1,tmdc_soc_angular}. Because of the crystal symmetry, this spin splitting is expected to have the form $\beta(k_\parallel)\cos(3 \theta)$ with out-of-plane polarization\cite{tmdc_soc1}. In Fig. \ref{fig:tbh_soc} (b), we compare the angular dependence of the spin splitting at $k_\parallel = K/2$ and $k_\parallel = K$ in TBH + SOC (red dots) with the fitted $\cos(3 \theta)$ formula (black line). This fitted formula holds well over the BZ and the spin splitting for the BZ can be determined with the radial data in Fig. \ref{fig:tbh_soc} (a) multiplied by the angular dependence, that is, the $\cos(3 \theta)$ term. The radial part at small $k$ rises as $k^3$, consistent with the dominant spin splitting term $(k_x+ik_y)^3+(k_x-ik_y)^3$ from the crystal symmetry. A more accurate SOC effect will be presented in connection to the k $\cdot$ p hamiltonians at the K$_\pm$ points which are tailored for low-energy physics with higher accuracy.

\section{\label{sec:level1}Interlayer Pair Coupling}

\subsection{\label{sec:level2}$p$-$p$ Interlayer Coupling}

\begin{figure}
\centering
\includegraphics[width=0.45\textwidth]{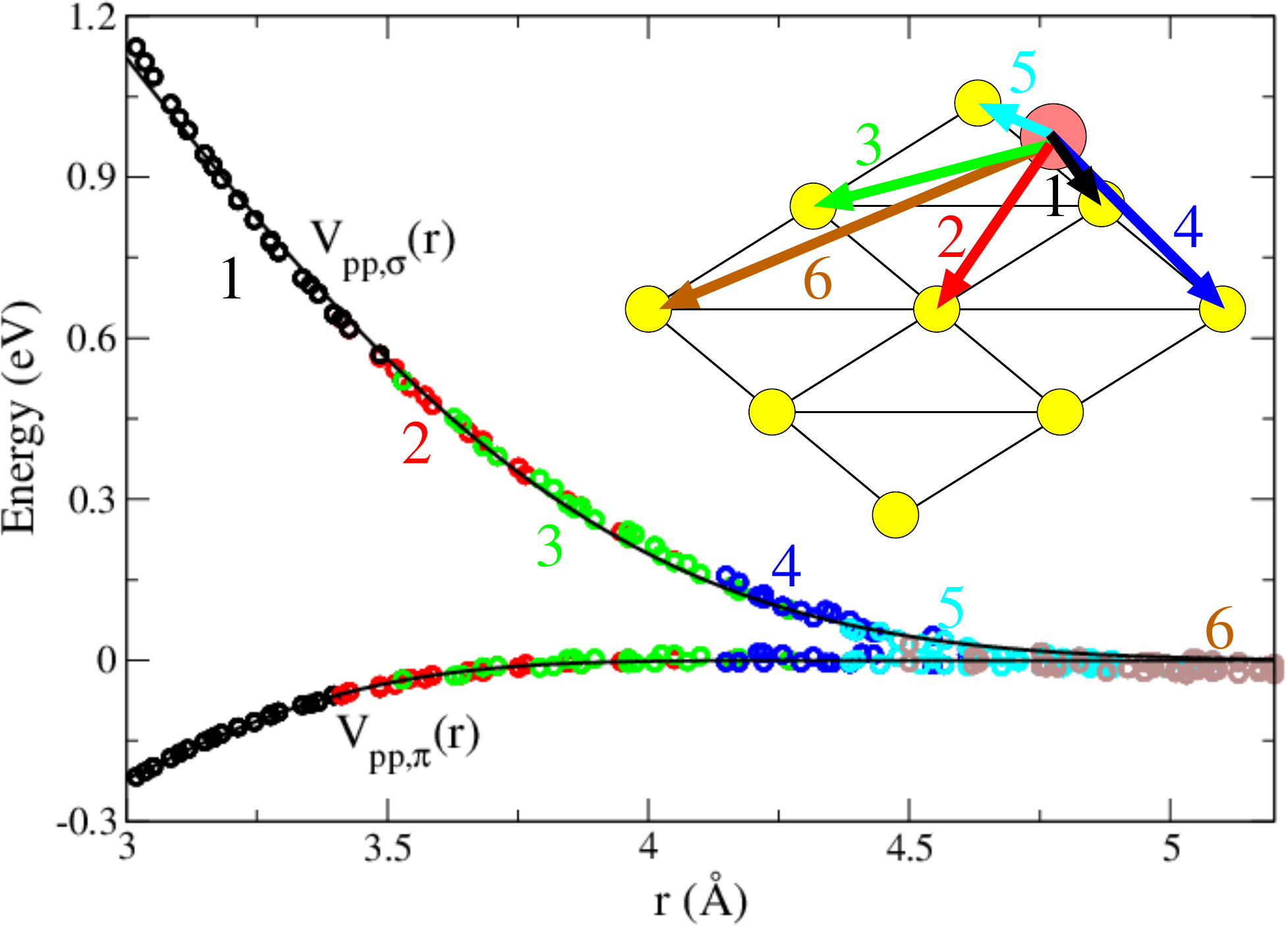}
\caption{$V_{pp,\sigma}(r)$ (upper) and $V_{pp,\pi}(r)$ (lower) for interlayer pairs as function of the pair distance $r$ in a MoS$_2$ bilayer unit.  Black, red, green, blue, cyan, brown are data points from 1st, 2nd, 3rd, 4th, 5th, 6th nearest distances. The solid black lines going through the points are the fits as obtained from Eq. \ref{eqn:tmdc_interlayer}. The inset is the perspective view for the interlayer coupling between one of the S atom on the top unit (orange) to all other S atoms on the bottom unit (yellow).}
\label{fig:interlayer}
\end{figure}

In multilayer structures or the heterostructures of several different 2D layers, attractive forces of van der Waals type bind the individual layers together. The resulting electronic states show interesting new features due to the interlayer interactions. In this section we investigate the induced changes in electronic band structure and the interaction hamiltonian between layers of the same kind. We stress that the many-body physics of the van der Waals interactions is not included in the electronic band structure. In principle, this many-body effect should be present in the exchange-correlation functional, but is not taken explicitly into consideration here. The van der Waals force affects mostly the crystal structure and the equilibrium vertical separation of the layers. The 2H crystal structure is shown in Fig. \ref{fig:TMDC_crystal} (b) and the equilibrium distance of layers along the  \textit{c}-axis is taken from experimental values.

The dominant interaction contribution to the electronic band structure comes from orbital hybridization between adjacent X layers. These are the \textit{p}-\textit{p} hopping terms between X atoms, with $p_z$-$p_z$ coupling being the dominant contribution due to the orbital orientation. We will show that the two-center Slater-Koster approximation\cite{slater} describes interlayer coupling well and transferable empirical interactions can be obtained for the strength of $\sigma$ and $\pi$ bonds as functions of the pairing distance across the spacing between layers.

To extract the interlayer interaction, the full-range Wannier tight-binding hamiltonian is constructed for a 2H bilayer TMDC unit as in the monolayer case\cite{graphene_bilayer_mlwf}. The initial orbital projections are the same atomic orbitals and the Hilbert space twice as large. The hamiltonian can be separated into two parts: the first corresponds to the coupling within the same monolayer unit and it is in diagonal blocks of each layer, while the second, $\hat{H}^{\rm (2L)}_{\rm int}$, includes the interlayer hopping matrix elements, $t^{\rm (LL)}_{p'_i,p_j}$:

\begin{equation}
\label{eqn:ham_int}
\hat{H}^{\rm (2L)}_{\rm int} = \sum_{p'_i,\vec{r}_2,p_j,\vec{r}_1} \hat{\phi}^{\dagger}_{2,p'_i} (\vec{r}_2) t^{\rm (LL)}_{p'_i,p_j} (\vec{r}_2-\vec{r}_1) \hat{\phi}_{1,p_j} (\vec{r}_1) + {\rm h.c.}
\end{equation}

For each interlayer pair of X atoms, nine hopping parameters are needed to describe the interactions between three \textit{p} orbitals on each, denoted as:

\begin{equation}
t^{\rm (LL)}_{p'_i,p_j}(\vec{r}_2-\vec{r}_1)=<\phi_{2,p'_i}(\vec{r}_2)|\hat{H}^{\rm (2L)}_{\rm int}|\phi_{1,p_j}(\vec{r}_1)>
\end{equation} Within the two-center Slater-Koster approximation\cite{slater}, these nine \textit{p}-\textit{p} coupling parameters are reducible and can be simplified to two independent parameters, $V_{pp,\sigma}$ and $V_{pp,\pi}$, corresponding the $\sigma$ and $\pi$ bonds. The nine hopping terms for each pair reconstructed from the $V_{pp,\sigma}$ and $V_{pp,\pi}$ representation are given by:

\begin{equation}
\label{eqn:two_center_bond}
t^{\rm (LL)}_{p'_i,p_j}(\vec{r})=\Big( V_{pp,\sigma}(r)-V_{pp,\pi}(r) \Big) \frac{r_i r_j}{r^2}+V_{pp,\pi}(r) \delta_{i,j} \\
\end{equation}
with $r=|\vec{r}|$. The inverse relations are used to extract $V_{pp,\sigma}(r)$ and $V_{pp,\pi}(r)$ from these nine hopping parameters for each interlayer pair:

\begin{equation}
\begin{split}
&V_{pp,\pi}(\vec{r})=\frac{1}{2}\sum_i t^{\rm (LL)}_{p'_i,p_i}(\vec{r}) -\frac{1}{2}\sum_{i,j} t^{\rm (LL)}_{p'_i,p_j}(\vec{r}) \frac{r_i r_j}{r^2} \\
&V_{pp,\sigma}(\vec{r})=\sum_{i,j} t^{\rm (LL)}_{p'_i,p_j}(\vec{r}) \frac{r_i r_j}{r^2} 
\end{split}
\end{equation}

For a bilayer configuration, multiple pairs of interlayer distances can be identified starting from the dominant contribution of the shortest-distance pair. These are indicated by the arrows in different colors in the inset of Fig. \ref{fig:interlayer}. 



\subsection{\label{sec:level2}Universal Form for Interlayer Coupling}

A single fixed configuration of 2H bilayer TMDC unit gives pairs with distances at only a discrete set of values. A spatial translation or a twist in the crystal orientation for each layer unit can exist in the real heterostructure. To study the interlayer coupling in a more general crystal structure, we need to map out $V_{pp,\sigma}$ and $V_{pp,\pi}$ as functions of the pair distance $r$ treated as a continuous variable by varying the conventional 2H bilayer structure. We use a horizontal translation of the top unit relative to the bottom unit of the conventional 2H bilayer to obtain interactions as the distance of the pair is varied continuously. The vertical distance between the layers, $c/2$, is kept fixed when the relative translation is applied. Since there is no rotation between the two monolayer units, the primitive cell is the same with displaced atoms at the top layer. The vector for the translation is spanned by $l_1 \vec{a}_1+l_2 \vec{a}_2$ with $l_i$ between 0 and 1. For each crystal configuration that includes a translation we extract the sets of interlayer pairs and the $V_{pp,\sigma}$ and $V_{pp,\pi}$ values associated with them. As shown in Fig. \ref{fig:interlayer}, $V_{pp,\sigma}$ and $V_{pp,\pi}$ values from the interlayer pairs in various crystal configurations collapse on two single curves respectively. This demonstrates the validity of the two-center Slater-Koster approximation and the pair distance $r$ as the only variable needed to parametrize the interlayer pair interaction. We use an exponential form to fit $V_{pp,b}$ as functions of $r$ where $b$=$\sigma$, $\pi$.\cite{tb_fitting}

\begin{equation}
\label{eqn:tmdc_interlayer}
V_{pp,b}(r)=\nu_b \exp( -(r/R_b)^{\eta_b})
\end{equation} This interlayer interaction depends only on the species of X atoms, since M atoms are well inside the monolayer unit and their orbitals do not contribute directly to the interlayer interactions. We give the values of these parameters in Table \ref{table:interlayer_fitting} for S-S and Se-Se interlayer interactions.

\begin{table}[ht!]
\caption{Interlayer $V_{pp,\sigma}$ and $V_{pp,\pi}$ parameters.} 
\label{table:interlayer_fitting}
\centering
\vspace{1 mm}
\begin{tabular}{|c|c|c|c|c|c|c|}
\hline\hline
Parameters & $V_{pp,\sigma}^{\rm S-S}$ & $V_{pp,\pi}^{\rm S-S}$ & $V_{pp,\sigma}^{\rm Se-Se}$ & $V_{pp,\pi}^{\rm Se-Se}$ \\
\hline
$\nu$ (eV) & 2.627 & $-0.708$ & 2.559 & $-1.006$ \\
\hline
$R$(\AA) & 3.128  & 2.923 & 3.337 &  2.927 \\
\hline
$\eta$ & 3.859 & 5.724 & 4.114 & 5.185 \\
\hline
\end{tabular}
\end{table}


The validity of the two-center approximation can be attributed to the weak coupling of the two units in the bilayer structure. The Wannier functions of $p$ orbitals at X atoms show only small crystal field distortion. That is, the atomic X orbitals are insensitive to the local crystal environment, or the crystal orientation. The interlayer coupling parameters do not require an angular dependence as established by the results shown in Fig. \ref{fig:interlayer}. With this azimuthal symmetry, the pair distance information is enough to determine the interaction strength, irrespective of the pair orientation. This also means that the interlayer interaction will have the same form when the two units are rotated by an arbitrary angle relative to each other, which makes our TBH scheme applicable to heterostructure that involve a twist between layers.

The azimuthal symmetry of atomic orbitals that participate in the interlayer coupling is not generally expected due to the crystal field distortion of atomic orbitals from the presence of neighboring orbitals. One salient case is the tight-binding hamiltonian for bilayer graphene\cite{graphene_rev} in Bernal stacking. Two types of pairs, $\gamma_3$ and $\gamma_4$, share the same pair distance but have very different hopping strength. In general, the crystal field introduces higher angular momentum mixing to atomic orbitals. This gives rise to additional orientational dependence for interlayer interactions beyond the simple interlayer pair distance. A more general interaction model can be constructed to account for both pair distance and orientation, which are crucial for modeling generic two-dimensional layered materials.

\subsection{\label{sec:level2}Application to 2H Bilayer and Bulk}
\begin{figure}
\centering
\includegraphics[width=0.48\textwidth]{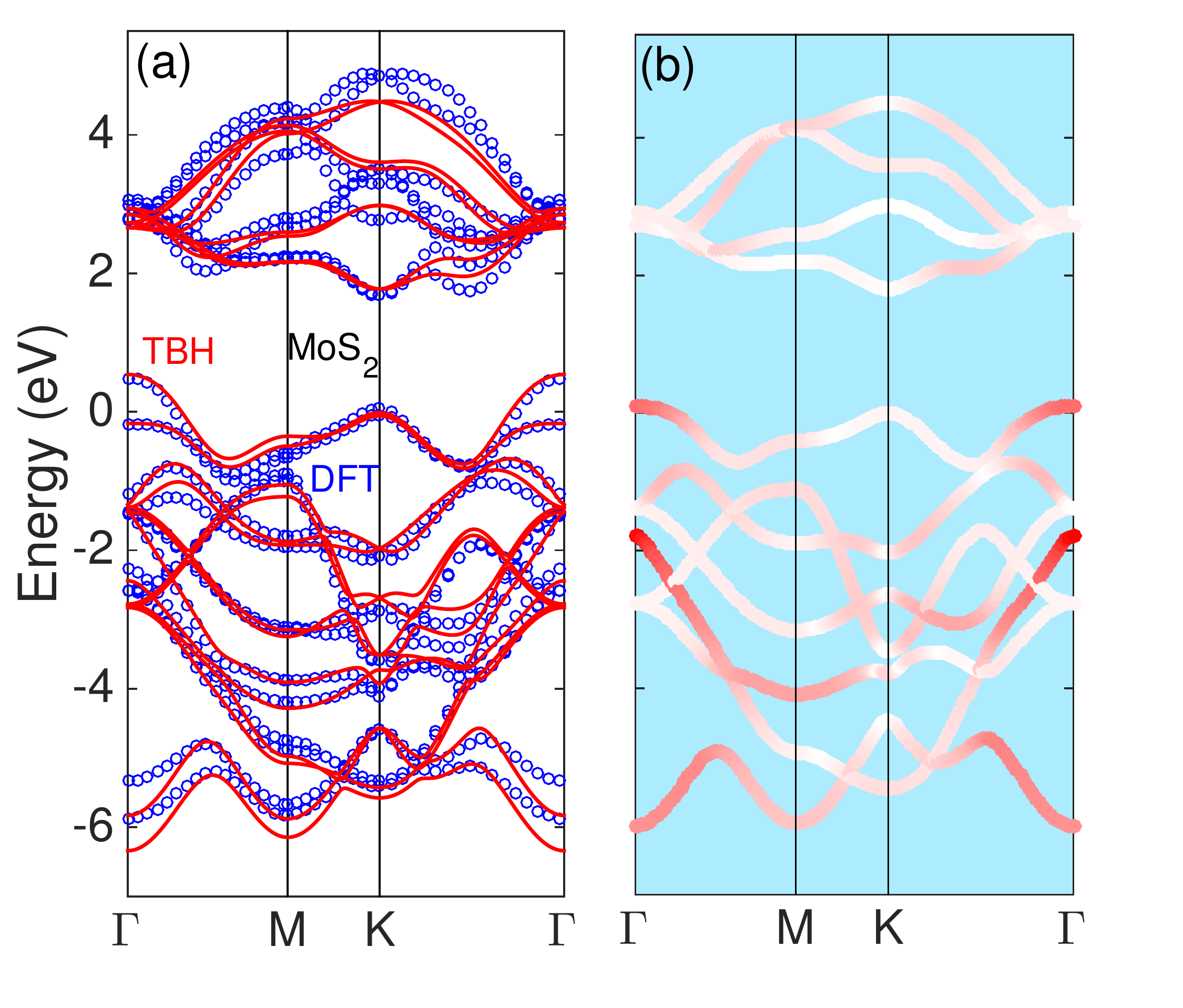}
\caption{(a)TBH band structure (red lines) along $\Gamma$-M-K-$\Gamma$ and the comparison with DFT results (blue circles) for a MoS$_2$ bilayer in the 2H structure. (b) The overlap matrix element strength between two layers with the interlayer coupling hamiltonian that leads to splitting of single-layer states, on a scale from red (strongest) to white (weakest).}
\label{fig:TMDC_2H}
\end{figure}

To test the interlayer interaction hamiltonian and compare with the full DFT calculation, we construct the tight-binding hamiltonian in Fourier space for a MoS$_2$ 2H bilayer as follows: 

\begin{equation}
\begin{split}
\label{eqn:ham_so}
\hat{H}^{\rm (2L)}&=\sum_k [\hat{\phi}^{\dagger}_{1}(k) H^{\rm (1L)}_{1}(\vec{k}) \hat{\phi}_{1} (k) +\hat{\phi}^{\dagger}_{2}(k) H'^{\rm (1L)}_{2}(\vec{k}) \hat{\phi}_{2} (k) \\
&+ \hat{\phi}^{\dagger}_{2}(k) V^{\rm (LL)}_{\rm int}(\vec{k}) \hat{\phi}_{1} (k) + h.c.]
\end{split}
\end{equation} Within each monolayer unit, $H^{\rm (TBH)}$ is used for the hamiltonian, $H^{\rm (1L)}$. The relative rotation for the second monolayer unit, $H'^{\rm (1L)}_2$, can be taken into account by either a rotation in the hamiltonian basis or xz mirror operation\cite{mos2_tb1}. For the interlayer hopping $V^{\rm (LL)}_{\rm int}$, all interlayer pairs with distance less than 5\AA $ $ are included. Due to the extension of $d_{z^2}$ orbitals into the interlayer region, additional interlayer coupling between $d_{z^2}$ and $p_z$ can be included to improve the accuracy of the highest valence bands at $\Gamma$ which are composed of $d_{z^2}$ and $p_2$ orbitals. These interlayer couplings are 60 meV for the nearest X-M and 26 meV for the second nearest X-M pairs. We show in Fig. \ref{fig:TMDC_2H} (a) the band structure with this bilayer tight-binding hamiltonian (red lines), which is in good agreement with the full DFT results (blue circles).

To gain insights of the effect of interlayer interactions, we can treat the interlayer couplings $\hat{V}^{\rm (LL)}$ in the context of a perturbative hamiltonian. When this perturbation is set to zero, the bands for the bilayer unit are doubly degenerate. For a generic $k$ point, we choose the eigenstates, $\varphi_1(\vec{k})$ and $\varphi_2(\vec{k})$ to exist on the individual layers respectively. The degenerate subspace is spanned by these two states. When the interlayer coupling $\hat{V}^{\rm (LL)}$ is turned on, it mixes the two states by introducing off-diagonal matrix elements, $<\varphi_1(\vec{k})|\hat{V}^{\rm (LL)}(\vec{k})|\varphi_2(\vec{k})>$. In Fig. \ref{fig:TMDC_2H}(b), the MoS$_2$ monolayer bands are plotted on a color scale that represents the magnitude of the interlayer off-diagonal matrix element with red representing the largest values and white the smallest values. The splitting between two degenerate states is proportional to this overlap matrix element. The strongest coupling and splitting comes from the bands with dominant $p_z$ character near the $\Gamma$ point. This hybridization drives the direct band gap into an indirect one when more than one layer is included in the structure. 

The interaction does not always guarantee the off-diagonal mixing between these two states due to symmetry selection rules. The symmetry and the orbital character (see Table \ref{table:orbital_sym}) can be used together to explain the degeneracies at K for the 2H bilayer. The degeneracy occurs when the two states, $\varphi_1(\rm K)$ and $\varphi_2(\rm K)$, at the same energy transform differently under $\mathcal{R}_3$ symmetry. As a result, the off-diagonal matrix elements of interlayer coupling vanish in this case. Due to the restoration of inversion symmetry in the 2H bilayer, the bands are doubly degenerate with two spins when spin-orbit coupling is included.


\begin{figure}
\centering
\includegraphics[width=0.43\textwidth]{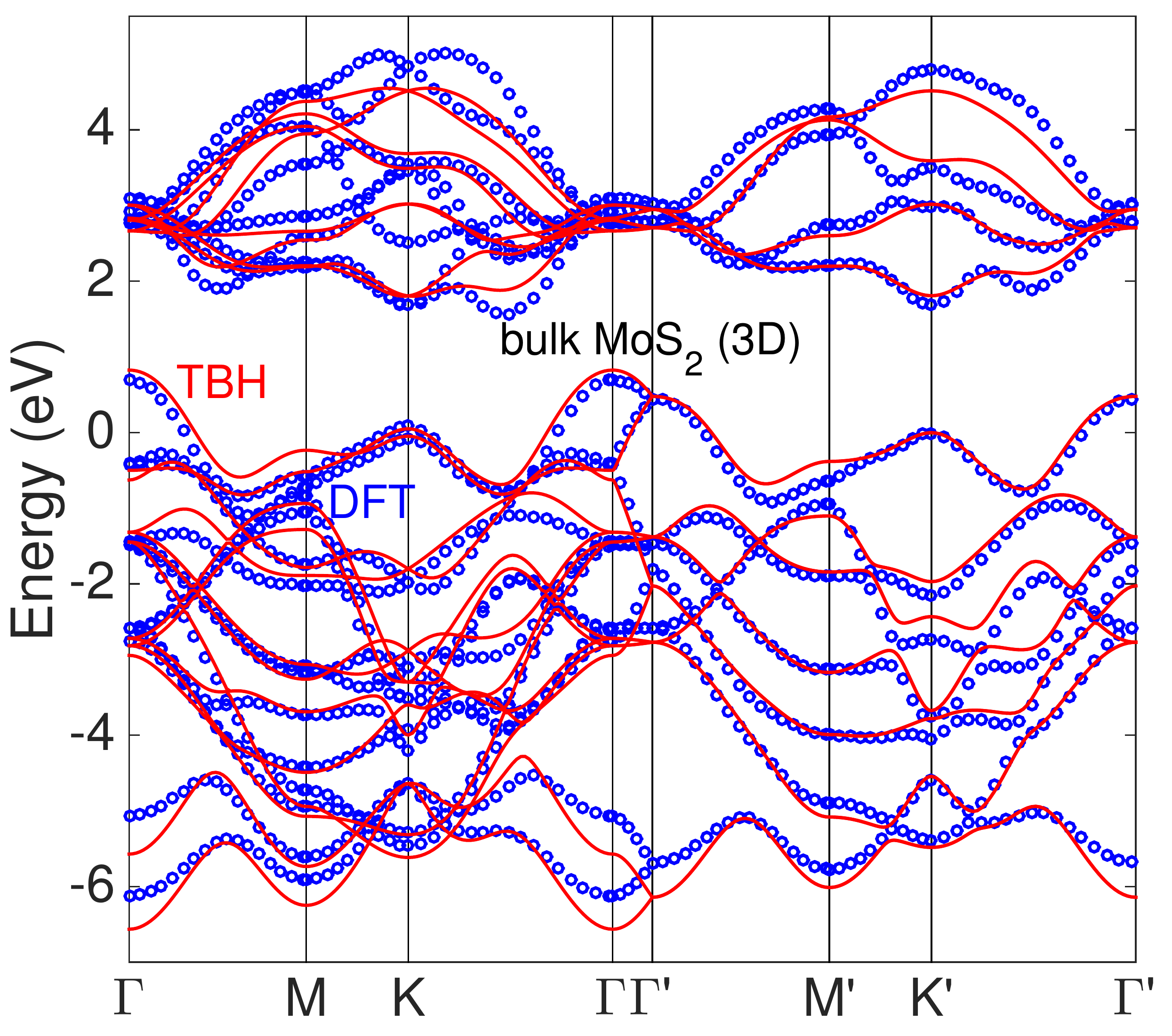}
\caption{Band structure of the bulk, 3D crystal of MoS$_2$, in the 2H structure, as obtained from DFT calculations (blue circles) and from the TBH (red lines) derived from the bilayer system (including inter-layer coupling). The symbols $\Gamma$-M-K-$\Gamma$ correspond to $k_z = 0$, while those with prime correspond to $k_z=\pi/c$.}
\label{fig:TMDC_2H_bulk}
\end{figure}

For the bulk band structure in the 2H configuration, there will be additional interlayer hopping terms under periodic boundary conditions compared to the bilayer unit. To have the minimal change to the monolayer hamiltonian and to incorporate properly the $k_z$ dependence, we make the gauge choice that for the $z$ direction atoms in the same layer are treated as if they are on the same $z$ plane. All the $k_z$ dependence enters into the interlayer terms and effectively involves a shift by $c/2$ in the $z$ direction for the interlayer pair. This gauge choice makes it easier to write the hamiltonian within each monolayer unit and renders them the same as the case with $k_z=0$. The resulting bulk bands from the tight-binding hamiltonian (red lines) are shown in Fig. \ref{fig:TMDC_2H_bulk}; these results clearly show that the bulk band structure is in good agreement with the full DFT results (blue dots).

\section{\label{sec:level1}k $\cdot$ p Hamiltonian at Band Extrema}
For the low-energy excitations around the band gap edges, it is useful to have an accurate perturbative approach in the form of a k $\cdot$ p hamiltonian. We construct this type of hamiltonian by performing a series expansion of the FTBH. In contrast to the TBH which faithfully represents the symmetry, band structure and orbital character over the entire BZ, the k $\cdot$ p hamiltonian is constructed and optimized for small regions of the BZ with selected bands where the low-energy excitations reside. We perform this k $\cdot$ p expansion for: (1) the highest valence band at $\Gamma$; (2) the lowest conduction and the highest valence bands at K$_\pm$. The expansions are based on the hamiltonian of the spinless electrons of the single layer for each material. The corresponding correction terms for the spin-orbit coupling or the interlayer interaction in the bilayer structure can be introduced as additional terms. In general, the presence of these perturbation terms would renormalize the parameters for the unperturbed hamiltonian of the single layer. We include these higher order effects as needed, and keep only the necessary lowest order terms otherwise.

The derivation of the k $\cdot$ p hamiltonian from the series expansion of the FTBH, as opposed to fitting the band structure of $\textit{ab-initio}$ results\cite{tmdc_spin_valley,tmdc_kp}, allows us to keep the phase and the orbital information from the analysis of DFT and FTBH results. The spinless k $\cdot$ p hamiltonian of the single layer is expanded up to second order in $k$-space to capture the effective masses for the bands. There are two contributions for the second order terms, originating from the second order expansion of the hamiltonian within the subspace or the second order virtual transition terms. The latter comes from the effective terms by integrating out the irrelevant bands through a Schrieffer-Wolff transformation\cite{sch_wolff}. 

The basis refers to the Bloch bands at $\Gamma$ and K$_\pm$ and consists of hybrids of $p$ and $d$ orbitals as given in Table \ref{table:orbital_sym}. At the $\Gamma$ point, the valence band is of predominantly $d_{z^2}$ character. For the K$_\pm$ point, the conduction band is mostly of $d_{z^2}$ character while the valence band is mainly composed of $(d_{x^2-y^2}+i\tau d_{xy})/\sqrt{2}$ orbitals with $\tau =\pm 1$ for K$_\pm$. The gauge choice for the basis is to have as the real positive component either $d_{z^2}$ or $d_{x^2-y^2}$, depending on which one dominates. The group of the wave vector dictates the form for the hamiltonian under symmetry transformations\cite{group_theory_md}. Time reversal symmetry links the hamiltonian at K$_\pm$. In the following discussion, $\hat{\sigma}$ acts on the conduction and valence degrees of freedom at K and $\hat{\sigma}_x, \hat{\sigma}_y, \hat{\sigma}_z$ are the Pauli matrices. $\hat{s}$ acts on the spin subspace while $\hat{\mu}$ acts on the layer index when a bilayer is considered. The direct product of these operators and the Hilbert space is implicitly assumed when extra degrees of freedom and the perturbative terms are included.

We begin with the k $\cdot$ p spinless hamiltonian for a single layer. The zero energy point is the highest valence band at K and $\vec{k}=(k_x,k_y)$ is the relative crystal momentum from the expansion point. At the $\Gamma$ point, the valence band is an isotropic parabolic band due to $\mathcal{R}_3$ symmetry. The parameter $g_0$ is the relative energy shift of valence bands with respect to K and $a$ the lattice constant. These considerations lead to the following form:

\begin{equation}
\mathcal{H}^\Gamma(\vec{k})=g_0+g_1 a^2 |\vec{k}|^2
\end{equation} 

The structure of the k $\cdot$ p hamiltonian at K$_\pm$ is richer because both conduction and valence bands are involved. Time-reversal symmetry demands $H_{-\tau}(k_x,k_y)=H^*_{\tau}(-k_x,-k_y)$ between two valleys and the hamiltonian reads\cite{mos2_tb2}

\begin{equation}
\begin{split}
&\mathcal{H}^{\rm K}_\tau(\vec{k})=  \frac{f_0}{2} (1+\hat{\sigma}_z) + f_1 a (\tau k_x \hat{\sigma}_x +k_y \hat{\sigma}_y)   \\
& + a^2 |\vec{k}|^2 (f_2 +f_3 \hat{\sigma}_z) + f_4 a^2(k_x^2 \hat{\sigma}_x-k_y^2 \hat{\sigma}_x - 2\tau k_x k_y \hat{\sigma}_y)
\end{split}
\end{equation} In general, other off-diagonal forms which are consistent with the crystal symmetry can also appear in a different gauge choice of the basis. Up to linear order, this hamiltonian has the form of the massive Dirac equation with energy gap $f_0$. Expanding this 2 $\times$ 2 hamiltonian, the eigenvalues and the dispersion up to second order for the conduction and valence bands are approximately given by:

\begin{equation}
\label{eqn:kdotp_mass}
\begin{split}
\epsilon^c (\vec{k})  &\approx f_0+(f_2+f_3+\frac{f_1^2}{f_0}) a^2 |\vec{k}|^2  \\
\epsilon^v(\vec{k})  &\approx (f_2-f_3-\frac{f_1^2}{f_0}) a^2 |\vec{k}|^2 
\end{split}
\end{equation} The effective masses are obtained from the coefficients of the quadratic terms and are dominated by the massive Dirac terms $f_1^2/f_0$ while the $f_3$($f_2$) term gives additional particle-hole symmetric (asymmetric) effective mass contributions. 


\begin{figure}
\centering
\includegraphics[width=0.5\textwidth]{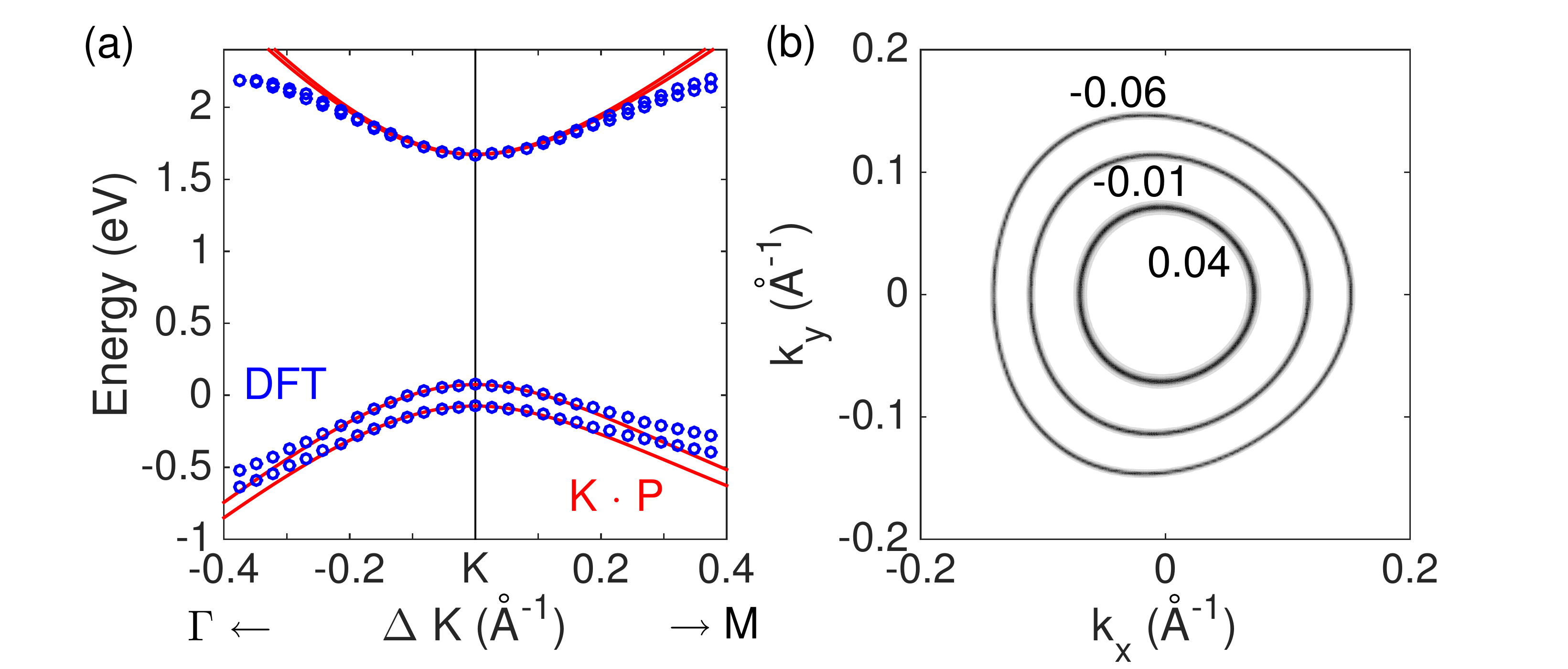}
\caption{(a) Comparison of the k $\cdot$ p hamiltonian (red lines) and the DFT results (blue dots) for the MoS$_2$ monolayer with spin-orbit coupling at K$_+$. (b) Energy contours in eV for the highest valence band centered at K$_+$. The deviation from spherical symmetry is due to the $f_4$ term.}
\label{fig:kdotp_plot}
\end{figure}

When spin is included, the Hilbert space is doubled and there will be additional spin-orbit coupling terms. At the $\Gamma$ point, the bands are doubly degenerate due to Kramers' theorem and there are no additional spin-orbit coupling terms to zeroth order. The spin splitting is non-zero when moving away from $\Gamma$ and it scales as $k^3$ (see the discussion for Fig. \ref{fig:tbh_soc}). At the K point, spin-orbit coupling splits the two spin states for both the conduction and valence bands. The lowest order four-dimensional spin-orbit coupling term is:

\begin{equation}
\Delta \mathcal{H}^{\rm K}_\tau=f_5 \tau (\frac{1-\hat{\sigma}_z}{2}) \hat{s}_z+f_6 \tau (\frac{1+\hat{\sigma}_z}{2}) \hat{s}_z
\end{equation} where  $f_5$ ($f_6$) represents the spin-orbit coupling strength for the valence (conduction) band. The small spin-splitting term $f_6$ for the conduction bands at K is more subtle from the two contributions in the perturbative expansion\cite{tmdc_soc2}. The first part is the negative contribution from the projection of the spin-orbit coupling operator within the conduction band subspace, while the second positive contribution is from the virtual coupling to other higher bands by the spin-orbit coupling. These two contributions tend to cancel out and result in the alternating signs in $f_6$ in MoX$_2$ and WX$_2$ as reported in the literature\cite{tmdc_soc2}. Due to time-reversal symmetry and broken inversion symmetry, the spin state flips when changing valley and the combination $\tau \hat{s}_z$ can be viewed as the effective coupling term between valley and spin. In the presence of spin splitting, the gap between bands is different and this will affect the dispersion compared to the spinless case. The dispersion for the bands with spin is given by the same expression as in Eq. (\ref{eqn:kdotp_mass}) with the gap $f_0$ replaced by the new spin-valley dependent gap $\Delta_{\tau s_z} = f_0+\tau s_z (f_6-f_5)$.

In Fig. \ref{fig:kdotp_plot}(a), the MoS$_2$ k $\cdot$ p hamiltonian is compared to the DFT bands around the K$_+$ point. In Fig. \ref{fig:kdotp_plot}(b), we plot the energy contours for the valence bands at several energies by computing the spectral function $\mathcal{A}(\vec{k},\omega)={-\rm Im}[{\rm Tr}[(\omega-\mathcal{H}+i\delta)^{-1}]]/\pi$ with a smearing $\delta = 2$ meV. The $f_4$ term gives the deviation of the energy contours from spherical symmetry due to the underlying three-fold crystal symmetry\cite{mos2_warping}. 


\begin{table}[ht!]
\centering
\label{table:kp_parm}
\caption{k $\cdot$ p hamiltonian parameters in eV with the lattice constant $a$ in \AA.} 
\vspace{3 mm}
\begin{tabular}{|c|c|c|c|c|c|}
\hline
 & {\bf MoS$_2$} & {\bf MoS$_2$ GW}  & {\bf MoSe$_2$} & {\bf WS$_2$} & {\bf WSe$_2$} \\
\hline
$a$ &  3.18  &  3.18 & 3.32  & 3.18 & 3.32 \\
\hline
\hline             
$g_0$ & $-0.0167$ & $-0.1161$ & $-0.2712$ & $-0.0648$ & $-0.3347$\\
\hline
$g_1$ & $-0.1173$ & $-0.1004$ & $-0.0642$ & $-0.1314$ & $-0.0680$\\
\hline
\hline
$g_2$ & $ 0.1563 $ & $-$ & $-0.0705 $ & $ 0.0940 $ & $ -0.1570 $ \\
\hline
$g_3$ & $ -0.2106 $ & $-$ & $ -0.1351 $ & $ -0.2365 $ & $ -0.1534 $ \\
\hline
$g_4$ & $ 0.1699 $ & $-$ & $ 0.1464 $ & $ 0.1869 $ & $ 0.1599$ \\
\hline
$g_5$ & $-0.3319 $ & $-$ & $-0.3352 $ & $-0.3272 $ & $-0.3205 $ \\
\hline
\hline
$f_0$ & $1.6735$ & $2.4826$ & $1.4415$ & $1.8126$ & $1.5455$\\
\hline
$f_1$ & $1.1518$ & $1.7381$ & $0.9560$ & $1.4073$ & $1.1894$\\
\hline
$f_2$ & $0.0744$ & $0.0957$ & $0.0494$ & $0.1551$ & $0.1184$\\
\hline
$f_3$ & $-0.0613$ & $-0.2917$ & $-0.0493$ & $-0.0175$ & $-0.0064$\\
\hline
$f_4$ & $-0.0780$ & $-0.1308$ & $-0.0654$ & $-0.0709$ & $-0.0627$ \\
\hline
\hline
$f_5$ & $0.0746$ & $-$ & $0.0929$ & $0.2153$ & $0.2335$  \\
\hline
$f_6$ & $-0.0015$ & $-$ & $-0.0106$ & $0.0148$  & $0.0180$  \\
\hline
$f_7$ & $-0.0417$ & $-$ & $-0.0508$ & $-0.0502$ & $-0.0612$ \\
\hline
\end{tabular}
\end{table}

In the 2H bilayer unit, the interlayer coupling introduces an off-diagonal term to mix the two layers and leads to splitting of bands that come from each layer. Certain complications arise from the difference in crystal orientation of the two monolayer units. The k $\cdot$ p hamiltonian for the bilayer unit at $\Gamma$ reads:

\begin{equation}
\mathcal{H}^{\Gamma,\rm 2H}(\vec{k}) =g_2 + a^2 |\vec{k}|^2  (g_3+g_4\hat{\mu_x}) + g_5 \hat{\mu}_x
\end{equation}

The interlayer coupling mixes the states on two layers by the $g_5$ term. The coupling also introduces a renormalized constant energy shift, $g_2$, for the central position of these bands, instead of $g_0$ in a single layer unit. This shift comes from the virtual hopping process to other bands which are integrated out. In this case, this is mostly from the hybridization between the top valence bands at $\Gamma$ and number four band below on the other layer with $p_0$ character (see Table \ref{table:orbital_sym}). The energy shift is positive from second order perturbation theory and is proportional to the square of the overlap matrix element and the inverse of the energy difference. The effective masses are renormalized as well in the presence of interlayer coupling and these effects are incorporated in the values of the parameters $g_3$ and $g_4$.

At the K point, the orbital character and k $\cdot$ p forms for these two layers are different and these can be accounted for by an xz mirror operation on the basis and the hamiltonian. Without the interlayer coupling, the hamiltonian for a spinless 2H bilayer is:

\begin{equation}
\begin{split}
&\mathcal{H}^{\rm K, 2H}_{\tau}(\vec{k}) = \frac{f_0}{2} (1+\hat{\sigma}_z)+ f_1 a (\tau k_x \hat{\sigma}_x +k_y \hat{\mu}_z  \hat{\sigma}_y)   \\
&+ a^2 |\vec{k}|^2  (f_2 +f_3 \hat{\sigma}_z) + f_4 a^2  (k_x^2 \hat{\sigma}_x-k_y^2 \hat{\sigma}_x - 2 \tau k_x k_y \hat{\mu}_z\hat{\sigma}_y) 
\end{split}
\end{equation} The form for spin-orbit coupling is:
\begin{equation}
\Delta \mathcal{H}^{\rm K, 2H}_\tau = f_5 \tau \hat{\mu}_z (\frac{1-\hat{\sigma}_z}{2}) \hat{s}_z +f_6 \tau \hat{\mu}_z (\frac{1+\hat{\sigma}_z}{2}) \hat{s}_z
\end{equation} and the interlayer coupling term is:
\begin{equation}
\Delta \mathcal{H}^{\rm K, 2H}=f_7  (\frac{1-\hat{\sigma}_z}{2})  \hat{\mu}_x
\end{equation}

We note that in spinless form, only the valence bands will split while the conduction bands remain doubly degenerate by symmetry. The numerical values of the coefficients for these k $\cdot$ p hamiltonians are given in Table \ref{table:kp_parm}, including GW results for spinless monolayer MoS$_2$.

\section{\label{sec:level1}Applications}
In what concerns applications of the hamiltonians we have derived here, optical transitions\cite{tmdc_spin_valley,graphene_phase} and Berry curvature effects\cite{berry_rev} are two important factors that control low-energy dynamics of electrons at different valleys. To incorporate interband coupling effects, we use the spinless eleven-band TBH to evaluate these physical quantities. The generalization to the spin case is straightforward. 

\subsection{\label{sec:level2}Interband Optical Transition}

Light couples to the Bloch bands through the gauge field. A heuristic way to derive the optical transition matrix is to start from the expansion of the tight-binding hamiltonian at an arbitrary $\vec{k}_0$, $\hat{H}(\vec{k}+\vec{k}_0) \approx \hat{H}(\vec{k}_0) + \vec{k} \cdot \partial \hat{H}(\vec{k})/\partial \vec{k}|_{\vec{k}_0}$. The gauge field $\vec{A}$ enters in the replacement $\vec{k} \rightarrow \vec{k}+e\vec{A}/\hbar$ with the electron charge $-e$. The interaction term with the gauge field is $\mathcal{P}(\vec{k}) = A_x \mathcal{P}_x(\vec{k}) +A_y \mathcal{P}_y(\vec{k})$ where $ \mathcal{P}_i(\vec{k})= (e/\hbar)\partial \hat{H}(\vec{k})/\partial k_i$ \cite{graphene_phase} and $\vec{A}$ is determined by the polarization of light. This is the current operator for the hopping terms. For $\sigma_\pm$ circularly polarized light, $\mathcal{P}_\pm(\vec{k}) = \mathcal{P}_x(\vec{k}) \pm i\mathcal{P}_y(\vec{k})$. The interband optical transition rate, obtained from Fermi's golden rule, between valence and conduction bands is determined by the matrix element of $\mathcal{P}(\vec{k})$ between these two Bloch wavefunctions $\psi_i$:

\begin{equation}
I(\omega) \sim \omega^{-2}\sum_{\vec{k},c,v} |<\psi_{c\vec{k}} | \mathcal{P}(\vec{k}) | \psi_{v\vec{k}}>|^2 \delta(\epsilon_c({\vec{k}})-\epsilon_v({\vec{k}})-\hbar \omega)
\end{equation} The crystal momentum $\vec{k}$ is the same for these two Bloch states in the optical limit and the $\omega^{-2}$ factor is from the conversion between $\vec{A}^2$ and the incoming light intensity $\vec{E}^2$.

\begin{figure}
\centering
\includegraphics[width=0.5\textwidth]{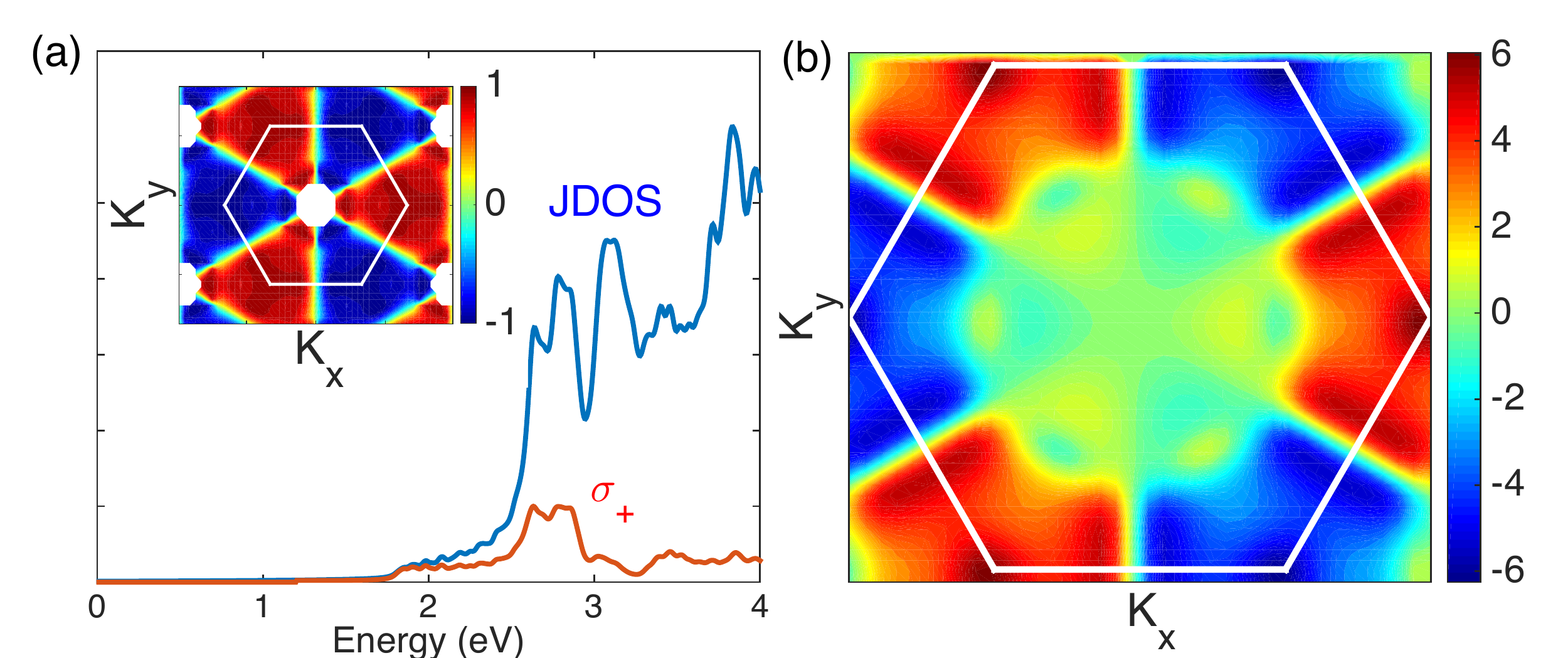}
\caption{MoS$_2$ TBH: (a) $\sigma_+$ absorption and comparison with joint density of states (JDOS);the inset shows the $k$-resolved degree of polarization $\eta(\vec{k})$. (b) Berry curvature in units of \AA$^2$.}
\label{fig:tbh_application}
\end{figure}

In Fig. \ref{fig:tbh_application}(a), we compute $I_+(\omega)$ numerically for the integrated $\sigma_+$ absorption over the entire Brillouin zone and compare with the joint density of states (JDOS), $\sum_{\vec{k},c,v} \delta(\epsilon_c({\vec{k}})-\epsilon_v({\vec{k}})-\hbar \omega)$, between conduction and valence bands. Light absorption starts at the band edge between the top valence band and lowest conduction band. Because of the broken inversion symmetry, the two valleys at K$_\pm$ respond differently to circularly polarized light\cite{dichroism_tmdc,valley_opts}. At K$_+$ the orbital character of the highest valence band and the lowest conduction band is $d_2$ and $d_0$ respectively. $\sigma_\pm$ polarized light adds angular momentum $\delta m_z=\pm 1$. Under $\mathcal{R}_3$ symmetry, only $\sigma_+$ can have a non-zero matrix element for the optical transition due to angular momentum conservation. To demonstrate the optical chiral selection rule in our hamiltonian, we compute the circular dichroism $\eta(\vec{k})$\cite{dichroism_tmdc}, given by:

\begin{equation}
\eta(\vec{k}) = \frac{|\mathcal{P}_+^{cv}(\vec{k})|^2-|\mathcal{P}_-^{cv}(\vec{k})|^2}{|\mathcal{P}_+^{cv}(\vec{k})|^2+|\mathcal{P}_-^{cv}(\vec{k})|^2}
\end{equation} This is the $k$-resolved quantity to distinguish $\sigma_+$ and $\sigma_-$ light absorption between the highest valence and the lowest conduction bands with the matrix element given by $\mathcal{P}_\pm^{cv}(\vec{k})=<\psi^c_{\vec{k}} | \mathcal{P}_\pm(\vec{k}) | \psi^v_{\vec{k}}>$. The photon energy is implicitly chosen to match the energy difference between the two bands at each $\vec{k}$ point. In the inset of Fig. \ref{fig:tbh_application}(a) we plot $\eta(\vec{k})$: at K$_\pm$, it is either $+1$ or $-1$ due to the optical chiral selection rule. Near the $\Gamma$ point, the lowest conduction band is an odd state and the optical transition from the highest valence band vanishes due to the symmetry selection rule. However, we find a strong degree of optical polarization over a large region near K$_\pm$ where the selection rule is not obeyed. In reality, the degree of polarization is degraded by inter-valley scattering which is beyond the scope of the present discussion.


\subsection{\label{sec:level2}Berry Curvature}
In the semi-classical equation of motion for a wave packet, transport properties are controlled by the band dispersion and the anomalous velocity contribution from the Berry curvature $\Omega_n(\vec{k})$\cite{berry_rev,layer_pseudospin,berry_top} defined as: 


\begin{equation}
\begin{split}
&\Omega_n(\vec{k}) = i \hat{z} \cdot ( \vec{\bigtriangledown}_{\vec{k}} u_{n\vec{k}}^*) \times (\vec{\bigtriangledown}_{\vec{k}} u_{n\vec{k}})  \\
&=-2 (\frac{\hbar}{e})^2 \sum_{n' \neq n}\frac{{\rm Im} <u_{n\vec{k}}| \mathcal{P}_x(\vec{k}) |u_{n'\vec{k}}><u_{n'\vec{k}}|\mathcal{P}_y(\vec{k})|u_{n\vec{k}}>}{(\epsilon_n(\vec{k})-\epsilon_{n'}(\vec{k}))^2}
\end{split}
\end{equation} The integral of the Berry curvature over the entire 2D Brillouin Zone manifold gives the integer Chern number, also known as the TKNN number\cite{tknn_chern}, for that band. In Quantum Hall systems the sum of these integers for the filled bands corresponds to the Hall conductivity. Though the integral is zero for a time-reversal invariant system, the Berry curvature in TMDC monolayers is non-zero at generic $k$ points due to broken inversion symmetry. Time-reversal symmetry dictates that $\Omega_n(-\vec{k})=-\Omega_n(\vec{k})$. Based on the TBH model we derived, the Berry curvature can be readily evaluated\cite{chern_num}. In Fig. \ref{fig:tbh_application}(b) we plot the Berry curvature for the top valence band of MoS$_2$. The curvatures at the K$_\pm$ valleys are opposite to each other which means that the charge carriers in the two valleys will drift differently under an applied external electric field. 


\section{\label{sec:level1}CONCLUSION}
In this work, we derived  \textit{ab-initio} tight-binding hamiltonians for TMDCs based on a Wannier transformation. The model hamiltonians retain faithfully the overlap matrix elements, the orbital information and the accuracy of DFT calculations on which they are based. The eleven-band TBH with first-neighbor and partial second-neighbor couplings captures the features of the hybridization of the atomic $p$ and $d$ orbitals and reflects the $\mathcal{M}_1$, $\mathcal{M}_2$ and $\mathcal{R}_3$ symmetries in a monolayer unit. The spin-orbit coupling enters as the on-site LS terms related to individual atoms.

For multiple stacked TMDC layers, we determine the interlayer couplings between $p$ orbitals on adjacent X layers. The two-center Slater-Koster approximation works well for modeling interlayer coupling, with hopping terms $V_{pp,\sigma}(r)$ and $V_{pp,\pi}(r)$ expressed through a simple yet transferable empirical function of the pair distance. This interlayer interaction term is an essential ingredient in understanding TMDC heterostructures when there is a twist in the relative orientations of adjacent layers or translations between monolayer units. It would be interesting to generalize this interaction to other two-dimensional layered materials such as graphene and hBN where the pair orientation dependence is needed, in addition to the pair distance.

Another way to utilize the FTBH is to perform a k $\cdot$ p expansion for the low-energy bands at $\Gamma$ and K$_\pm$. The second-order k $\cdot$ p hamiltonian respects the crystal symmetry and is constructed to investigate the effects of spin-orbit coupling and interlayer hopping in the bilayer unit.

As examples of applications of our TBH, we investigate the optical absorption and the Berry curvature of MoS$_2$. The tight-binding hamiltonians can form the basis for further theoretical investigations, many-body physics, and simulations for potential applications under external electric or magnetic fields in finite-size nano-structures\cite{bilayer_tmdc,qdot_magnetic}, in either monolayer or heterostructure forms.

\begin{acknowledgements}
We thank Bertrand Halperin, Philip Kim, Mitchell Luskin, Oscar Granas, Dennis Huang and Dmitry Vinichenko for useful discussions. This work was supported by the STC Center for Integrated Quantum Materials, NSF Grant No. DMR-1231319 and by ARO MURI Award W911NF-14-0247. We used Odyssey cluster of the FAS by Research Computing Group at Harvard University, and the Extreme Science and Engineering Discovery Environment (XSEDE), which is supported by NSF Grant No. ACI-1053575.
\end{acknowledgements}

\appendix

\section{\label{sec:level1}TMDC TB-Model Simplifications and Numerical Parameters}
The 86 parameters for all $\epsilon_i$, $t_{i,j}^{(1)}$, $t_{i,j}^{(2)}$, $t_{i,j}^{(3)}$, $t_{i,j}^{(4)}$ and $t_{i,j}^{(5)}$ can be simplified to 36 independent parameters, which include a subset of $\epsilon_i$, and all $t_{i,j}^{(1)}$, $t_{i,j}^{(5)}$, by symmetry considerations. Under $\mathcal{M}_2$ and $\mathcal{R}_3$ symmetries, the hoppings to the symmetrical positions are not mutually independent. Here we summarize the simplifications of the truncated tight-binding parameters and tabulate their numerical values in Table \ref{table:wan_tb_0} and \ref{table:SOC}. For the sets of indexes: ($\alpha=1,\beta=2$), ($\alpha=4,\beta=5,\gamma=3$), ($\alpha=7,\beta=8,\gamma=6$), ($\alpha=10,\beta=11,\gamma=9$) with the first superscript index corresponding to ($+$) and the second to ($-$), we have the following relations:

\begin{equation}
\begin{split}
\epsilon_\alpha&=\epsilon_\beta \\
t_{\alpha,\alpha}^{(2)}&=\frac{1}{4} t_{\alpha,\alpha}^{(1)}+\frac{3}{4} t_{\beta,\beta}^{(1)} \\
t_{\beta,\beta}^{(2)}&=\frac{3}{4} t_{\alpha,\alpha}^{(1)}+\frac{1}{4} t_{\beta,\beta}^{(1)} \\
t_{\gamma,\gamma}^{(2)}&=t_{\gamma,\gamma}^{(1)}\\
t_{\gamma,\beta}^{(2,3)}&=\pm\frac{\sqrt{3}}{2} t_{\gamma,\alpha}^{(1)}-\frac{1}{2} t_{\gamma,\beta}^{(1)} \\
t_{\alpha,\beta}^{(2,3)}&=\pm\frac{\sqrt{3}}{4}( t_{\alpha,\alpha}^{(1)}-t_{\beta,\beta}^{(1)})-t_{\alpha,\beta}^{(1)} \\
t_{\gamma,\alpha}^{(2,3)}&=\frac{1}{2}t_{\gamma,\alpha}^{(1)}\pm \frac{\sqrt{3}}{2}t_{\gamma,\beta}^{(1)}
\end{split}
\end{equation}

while for ($\alpha=1,\beta=2,\alpha'=4,\beta'=5,\gamma'=3$), ($\alpha=7,\beta=8,\alpha'=10,\beta'=11,\gamma'=9$), we have:

\begin{equation}
\begin{split}
t_{\alpha',\alpha}^{(4)}&=\frac{1}{4}t_{\alpha',\alpha}^{(5)}+\frac{3}{4}t_{\beta',\beta}^{(5)} \\
t_{\beta',\beta}^{(4)}&=\frac{3}{4}t_{\alpha',\alpha}^{(5)}+\frac{1}{4}t_{\beta',\beta}^{(5)} \\
t_{\beta',\alpha}^{(4)}=t_{\alpha',\beta}^{(4)}&=-\frac{\sqrt{3}}{4}t_{\alpha',\alpha}^{(5)}+\frac{\sqrt{3}}{4}t_{\beta',\beta}^{(5)} \\
t_{\gamma',\alpha}^{(4)}&=-\frac{\sqrt{3}}{2}t_{\gamma',\beta}^{(5)} \\
t_{\gamma',\beta}^{(4)}&=-\frac{1}{2}t_{\gamma',\beta}^{(5)}  \\
t_{9,6}^{(4)}=t_{9,6}^{(5)}, \; t_{10,6}^{(4)}&=\frac{-\sqrt{3}}{2}t_{11,6}^{(5)}, \; t_{11,6}^{(4)}=\frac{-1}{2}t_{11,6}^{(5)}
\end{split}
\end{equation}


\begin{table}[ht!]
\caption{Tight-binding independent parameters in units of eV for MoS$_2$, MoSe$_2$, WS$_2$, WSe$_2$ based on the DFT results. }
\label{table:wan_tb_0}
\centering
\vspace{1 mm}
\begin{tabular}{|c|c|c|c|c|}
\hline\hline
 & {\bf MoS$_2$} & {\bf MoSe$_2$} & {\bf WS$_2$} & {\bf WSe$_2$} \\
\hline
$\epsilon_1=\epsilon_2$ & $1.0688$ & $0.7819$ & $1.3754$ & $1.0349$ \\
\hline
$\epsilon_3$ & $-0.7755$ & $-0.6567$ & $-1.1278$ & $-0.9573$ \\
\hline
$\epsilon_4=\epsilon_5$ & $-1.2902$ & $-1.1726$ & $-1.5534$ & $-1.3937$ \\
\hline
$\epsilon_6$ & $-0.1380$ & $-0.2297$ & $-0.0393$ & $-0.1667$ \\
\hline
$\epsilon_7=\epsilon_8$ & $0.0874$ & $0.0149$ & $0.1984$ & $0.0984$ \\
\hline
$\epsilon_9$ & $-2.8949$ & $-2.9015$ & $-3.3706$ & $-3.3642$ \\
\hline
$\epsilon_{10}=\epsilon_{11}$ & $-1.9065$ & $-1.7806$ & $-2.3461$ & $-2.1820$ \\
\hline
$t_{1,1}^{(1)}$ & $-0.2069$ & $-0.1460$ & $-0.2011$ & $-0.1395$ \\
\hline
$t_{2,2}^{(1)}$ & $0.0323$ & $0.0177$ & $0.0263$ & $0.0129$ \\
\hline
$t_{3,3}^{(1)}$ & $-0.1739$ & $-0.2112$ & $-0.1749$ & $-0.2171$ \\
\hline
$t_{4,4}^{(1)}$ & $0.8651$ & $0.9638$ & $0.8726$ & $0.9763$ \\
\hline
$t_{5,5}^{(1)}$ & $-0.1872$ & $-0.1724$ & $-0.2187$ & $-0.1985$ \\
\hline
$t_{6,6}^{(1)}$ & $-0.2979$ & $-0.2636$ & $-0.3716$ & $-0.3330$ \\
\hline
$t_{7,7}^{(1)}$ & $0.2747$ & $0.2505$ & $0.3537$ & $0.3190$ \\
\hline
$t_{8,8}^{(1)}$ & $-0.5581$ & $-0.4734$ & $-0.6892$ & $-0.5837$ \\
\hline
$t_{9,9}^{(1)}$ & $-0.1916$ & $-0.2166$ & $-0.2112$ & $-0.2399$ \\
\hline
$t_{10,10}^{(1)}$ & $0.9122$ & $0.9911$ & $0.9673$ & $1.0470$ \\
\hline
$t_{11,11}^{(1)}$ & $0.0059$ & $-0.0036$ & $0.0143$ & $0.0029$ \\
\hline
$t_{3,5}^{(1)}$ & $-0.0679$ & $-0.0735$ & $-0.0818$ & $-0.0912$ \\
\hline
$t_{6,8}^{(1)}$ & $0.4096$ & $0.3520$ & $0.4896$ & $0.4233$ \\
\hline
$t_{9,11}^{(1)}$ & $0.0075$ & $0.0047$ & $-0.0315$ & $-0.0377$ \\
\hline
$t_{1,2}^{(1)}$ & $-0.2562$ & $-0.1912$ & $-0.3106$ & $-0.2321$ \\
\hline
$t_{3,4}^{(1)}$ & $-0.0995$ & $-0.0755$ & $-0.1105$ & $-0.0797$ \\
\hline
$t_{4,5}^{(1)}$ & $-0.0705$ & $-0.0680$ & $-0.0989$ & $-0.0920$ \\
\hline
$t_{6,7}^{(1)}$ & $-0.1145$ & $-0.0960$ & $-0.1467$ & $-0.1250$ \\
\hline
$t_{7,8}^{(1)}$ & $-0.2487$ & $-0.2012$ & $-0.3030$ & $-0.2456$ \\
\hline
$t_{9,10}^{(1)}$ & $0.1063$ & $0.1216$ & $0.1645$ & $0.1857$ \\
\hline
$t_{10,11}^{(1)}$ & $-0.0385$ & $-0.0394$ & $-0.1018$ & $-0.1027$ \\
\hline
$t_{4,1}^{(5)}$ & $-0.7883$ & $-0.6946$ & $-0.8855$ & $-0.7744$ \\
\hline
$t_{3,2}^{(5)}$ & $-1.3790$ & $-1.3258$ & $-1.4376$ & $-1.4014$ \\
\hline
$t_{5,2}^{(5)}$ & $2.1584$ & $1.9415$ & $2.3121$ & $2.0858$ \\
\hline
$t_{9,6}^{(5)}$ & $-0.8836$ & $-0.7720$ & $-1.0130$ & $-0.8998$ \\
\hline
$t_{11,6}^{(5)}$ & $-0.9402$ & $-0.8738$ & $-0.9878$ & $-0.9044$ \\
\hline
$t_{10,7}^{(5)}$ & $1.4114$ & $1.2677$ & $1.5629$ & $1.4030$ \\
\hline
$t_{9,8}^{(5)}$ & $-0.9535$ & $-0.8578$ & $-0.9491$ & $-0.8548$ \\
\hline
$t_{11,8}^{(5)}$ & $0.6517$ & $0.5545$ & $0.6718$ & $0.5711$ \\
\hline
$t_{9,6}^{(6)}$ & $-0.0686$ & $-0.0691$ & $-0.0659$ & $-0.0676$ \\
\hline
$t_{11,6}^{(6)}$ & $-0.1498$ & $-0.1553$ & $-0.1533$ & $-0.1608$ \\
\hline
$t_{9,8}^{(6)}$ & $-0.2205$ & $-0.2227$ & $-0.2618$ & $-0.2618$ \\
\hline
$t_{11,8}^{(6)}$ & $-0.2451$ & $-0.2154$ & $-0.2736$ & $-0.2424$ \\
\hline
\end{tabular}
\end{table}

\begin{table}[ht!]
\caption{Atomic spin-orbit coupling strength in TBH in units of eV/$\hbar^2$.} 
\label{table:SOC}
\centering
\vspace{1 mm}
\begin{tabular}{|c|c|c|c|c|}
\hline\hline
 & {\bf Mo} & {\bf W} & {\bf S} & {\bf Se}  \\
\hline
$\lambda_{\rm SO}^{\rm M/X} $ & $0.0836$ & $0.2874 $ & $0.0556 $ & $0.2470$  \\
\hline
\end{tabular}
\end{table}

\section{\label{sec:level1}TMDC GW Quasi-Particle Calculation}

\begin{figure}
\centering
\includegraphics[width=0.4\textwidth]{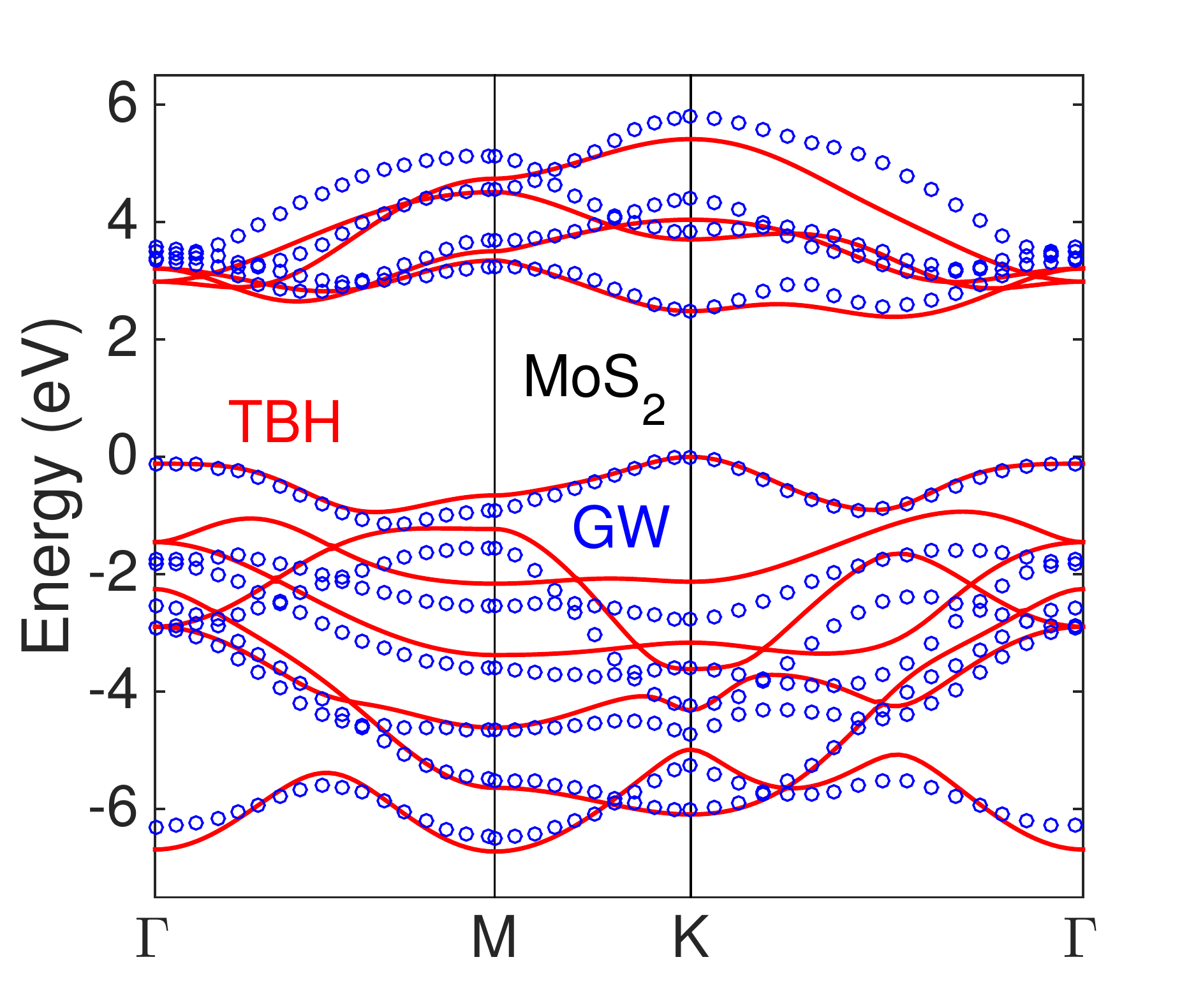}
\caption{Rescaled TBH hamiltonian (red lines) and its comparison with GW (blue circles) band structure results for a MoS$_2$ single layer unit.}
\label{fig:gw_tbh}
\end{figure}

We performed a GW calculation using the Quantum ESPRESSO\cite{qespresso} and BerkeleyGW\cite{bgw1,bgw2} code which corrects the band-gap by computing quasiparticle energies to obtain an improved model hamiltonian. The interactions between the ionic cores and electrons were modeled with normconserving pseudopotentials \cite{norm}. The exchange correlation energy between the electrons was treated within the Generalized Gradient Approximation (GGA) parameterized by Perdew, Burke and Ernzerhof (PBE) \cite{pbe}. The semi-core 4$d$, 4$p$, and 4$s$ states of Mo were taken as valence states for our DFT and GW calculations. All the integrations over the BZ were carried out over a 30 $\times$ 30 $\times$ 1 uniform mesh of $k$-points, and an energy cutoff of 1900 eV was used to truncate the plane wave basis used in representing Kohn-Sham wave functions. The self energy ($\sigma$) and the dielectric matrix ($\epsilon^{-1}_{\textbf{G,G'}}(\textbf{q})$) were calculated on a uniform mesh of 12 $\times$ 12 $\times$ 1 $k$-points in the BZ with a truncated Coulomb interaction. We employed an energy cutoff of 272  eV for the dielectric matrix, and 1750 bands to calculate $\epsilon^{-1}_{\textbf{G,G'}}(\textbf{q})$ and $\sigma$. The conduction and valence band quasiparticle energies are converged within 5 meV using the above parameters as discussed by Malone {\it et al.} \cite{bgw3}. 

The MoS$_2$ TBH based on GW can be constructed as in the DFT case. GW bands have similar shapes as those in DFT and a larger band gap between the groups of valence and conduction bands. This effect can be incorporated mostly by enlarged coupling constants between orbitals. Instead of tabulating a new column for GW-TBH parameters, we introduce atomic on-site energy shifts and scaling factors for the MoS$_2$ TBH parameters. Hopping terms in TBH are divided into four groups as different bonds in Fig. \ref{fig:TMDC_miniTB}. Each type of them are scaled by different scaling factors $z$, $t^{\rm GW}_{i,j}=zt^{\rm DFT}_{i,j}$, tabulated in Table \ref{table:MoS2_GW} under the condition to have the right GW band-gap, and the right values for the highest valence band energy at $\Gamma$. The GW-TBH is compared with the GW calculation in Fig. \ref{fig:gw_tbh}, showing good agreement between the two.

\begin{table}[ht!]
\caption{On-site energy shifts in eV and scaling factors for MoS$_2$ GW-TBH based on DFT-TBH.} 
\label{table:MoS2_GW}
\centering
\vspace{1 mm}
\begin{tabular}{|c|c|c|c|c|c|c|}
\hline\hline
& $\Delta \epsilon_{\rm M}$ & $\Delta \epsilon_{\rm X}$ & 1st M-M & 1st X-X  & 1st X-M & 2nd X-M  \\
\hline
{\bf MoS$_2$} & 0.3624 & -0.2512 & 1.4209 & 1.1738 & 1.0773 & 1.1871 \\
\hline
\end{tabular}
\end{table}

\end{document}